\documentclass[]{raa}            
\usepackage{graphicx,times}
\usepackage{natbib}

\providecommand{\bm}[1]{\mbox{\boldmath$#1$\unboldmath}}

\begin{document}

   \title{Fractal dimension and thermodynamic fluctuation properties of IDV light curves
}

\volnopage{ {\bf 20xx} Vol.\ {\bf 9} No. {\bf XX}, 000--000}
   \setcounter{page}{1}

\author{C. S. Leung
      \inst{1,2}
   \and J. Y. Wei
      \inst{1}
      \and A. K. H. Kong
      \inst{3}
   \and Z. Kov\'acs
      \inst{4}
   \and T. Harko
   \inst{4}
   }

\institute{National Astronomical Observatories, Chinese Academy of Sciences, 20A Datun Road,
   Chaoyang District, Beijing, China; {\it astrosinghk@yahoo.com.hk}\\
        \and
        Graduate School of Chinese Academy of Sciences, Beijing 100049, China
        \and
        Institute of Astronomy and Department of Physics, National Tsing Hua University,
Hsinchu, Taiwan
\and
      Department of Physics and Center for Theoretical and Computational Physics,
The University of Hong Kong, Pok Fu Lam Road, Hong Kong, Hong Kong SAR, P. R. China}

\vs \no
   {\small Received [year] [month] [day]; accepted [year] [month] [day] }

\abstract{Fractals are a basic tool to phenomenologically describe natural objects having a high degree of temporal or spatial variability. From a physical point of view the fractal properties of natural systems can also be interpreted by using the standard formalism of thermodynamical fluctuations. In the present paper we introduce and analyze the fractal dimension of the Intra - Day- Variability (IDV) light-curves of the BL Lac objects, in the optical, radio and X-ray bands, respectively. A general description of the fluctuation spectrum of these systems based on general thermodynamical principles is also proposed. Based on the general fractal properties of a given physical system, we also introduce the predictability index for the IDV light curves. We have explicitly determined the fractal dimension for the R-band observations of five blazars, as well as for the radio band observations of the compact extragalactic radio source J 1128+5925, and of several X-ray sources. Our results show that the fractal dimension of the optical and X-ray observations indicates an almost pure "Brownian noise" (random walk) spectrum, with a very low predictability index, while in the radio band the predictability index is much higher. We have also studied the spectral properties of the IDV light curves, and we have shown that their spectral index is very closely correlated with the corresponding fractal dimension.
\keywords{galaxies: quasars: emission lines: instabilities: gravitation}
}

   \authorrunning{C. S. Leung, J.~Y. Wei, A. K. H. Kong, Z.~Kov\'acs \& T. Harko}            
   \titlerunning{Fractal properties of IDV light curves }  
   \maketitle


%
%
\section{Introduction}           
\label{intro}

Fractals are a basic tool to phenomenologically describe natural
objects. According to \cite{Man}, a fractal is  a
set whose Hausdorff dimension is not an integer. The fractal dimension is one possible
parameter that characterizes chaotic systems, and the analysis of time series is one of the most common
means to find the fractal dimension from observable quantities. Fractals are also very useful for the analysis of the waveforms, a term that describes the shape of a wave, usually drawn as instantaneous values of a
variable quantity versus time. The fractal analysis of waveforms was introduced by \cite{Katz}, who considered the possibility that the complexity of a waveform may be represented by its fractal dimension. \cite{Katz} proposed that the fractal dimension can be measured empirically by sampling the waveform at $N$ points evenly spaced on the abscissa, which discretizes the waveform into $N'=N-1$ segments. Then the fractal dimension can be obtained as $D_F=\log N'/\left[\log N'+\log\left(d/L\right)\right]$, where $d=\max\left[{\rm dist}(i,j)\right]$ is the planar extent of the curve, and $L=\sum _{i=0}^{N'}{\rm dist}(i,i+1)$ is the length of the curve, where "max" stands for the maximum dist$(i, j)$, the distance between points $i$ and $j$ of the curve. An alternative and more efficient algorithm for the computation of the fractal dimension was proposed by \cite{Sevcik, Sevcik1}, which we will consider in the following.

The fractal analysis has been extensively used in astronomy and astrophysics for the study of the fractal properties of gamma ray bursts (\citealt{Sh}), the critical properties of spherically symmetric accretion in a fractal medium (\citealt{Ro}),  for  the analysis of fractal structures in the photospheric and the coronal magnetic field of the Sun (\citealt{Io,Di}), as a measure of the scale of homogeneity (\citealt{Ya}), for the study of the star forming regions (\citealt{Sa}), and for the analysis of the dark matter and gas distributions in the Mare-Nostrum universe (\citealt{Ga}), respectively. The quasar distribution on the celestial sphere is characterized by power laws as well with correlation dimension value equal from 1.49 to 1.58 for different redshift layers in the same range (\citealt{Roz1}). A fractal cosmological model, which accounts for the observable fractal properties of the large-scale structure of the Universe, was discussed in \cite{Roz2}.

Many classes of Active Galactic Nuclei have variability properties. The time scale of variability can be classified as yearly, monthly or daily. If the sources vary within one night or day, the time variation is called "Intra-day Variability", or simply IDV. There is one main method for confirming if the object is with IDV in the differential light curve, or not. This is the so-called statistical method. In this method one checks if the variability amplitude is greater than 3 times the sigma value, or not. The best criteria is however  5 times of sigma criteria (\citealt{Fan}). Another method, based on the so called C-parameter, was introduced by \cite{Rom}. The criteria for confirming the object is with IDV is the "C" value greater than or equal to 2.576. If this is the case, the IDV variability is confirmed at 99\% confidence level. Furthermore, the object is said to be variable if the "C" value is greater than or equal to 2.576 at two different bands. The IDV can be checked by the formula proposed by \cite{Heidt}.  For checking the IDV for the source in radio band a statistical parameter called modulation index is used. Usually, for the IDV light curve of blazars there is no significant pattern and period. \cite{Rol} proposed that the formation and the rotation of a warp in the inner part of the accretion disk can lead to some perturbations of the beam that finally produce the observed IDV.

It is the purpose of the present paper to analyze the IDV light curves by using their fractal dimension properties, and to try to interpret the actual value of the fractal dimension by using the general formalism of thermodynamic fluctuations. The actual IDV signal is always fluctuating. Profiles of IDV signals are also remarkably varied. The possible mechanisms for generating IDV emissions could also be different. The fractal dimension could be provide a hint into the physical processes that could trigger the IDV processes. In the present paper we have explicitly determined the fractal dimension for the R-band optical emission of five blasars (\citealt{Gu}) and for the radio band emission of the compact extragalactic radio source J 1128+5925 (\citealt{Gab}). The fractal dimension allows us the introduction of the predictability index, which represents a powerful indicator of the nature of the IDV signal. The general fractal properties of the IDV signals can be interpreted in the framework of the general theory of thermodynamic fluctuations for self-gravitating systems. The fractal dimension of the optical observations indicates an almost pure "Brownian noise" (random walk) spectrum, with a very low predictability index, while in the radio band the predictability index is much higher. We have also studied the spectral properties of the IDV light curves by using the periodogram method, and we have shown that their spectral index is very closely correlated with the corresponding fractal dimension.

The present paper is organized as follows. In Section~\ref{sect1} we consider the theory of the thermodynamic fluctuations of the self-gravitating systems, and obtain the general theoretical predictions for the fractal dimension of the fluctuation spectrum. The methods for the calculation of the fractal dimension and of the power spectrum are presented in Section~\ref{sect2}. The fractal dimensions and the power spectra of several IDV signals in optical, radio and X-ray bands are obtained in Section~\ref{sect3}. The correlation between the fractal dimension and the spectral index is also discussed in detail. We discuss and conclude our results in Section~\ref{sect4}.


\section{Fractal properties of fluctuating self-gravitating systems}\label{sect1}

In the following we consider a system of self-gravitating particles (an astrophysical accretion disk, for example), embedded in a heat bath (an external medium) of gravitationally non-interacting particles. The system interacts with the surrounding bath through a friction force, due to which the particles in Brownian motion lose energy to
the medium,  and simultaneously gains energy from the random kicks of the thermal bath, which can be described by a stochastic force (for a description of the dynamics of the system via a Langevin equation see \cite{Leung}). As a result of the interaction between the system and the heat bath, the thermodynamical quantities of the system fluctuate. In the following we consider the standard thermodynamic fluctuation formalism (\citealt{LaLi, Kan}) as applied to self-gravitating systems, and we derive the time dependence of the fluctuation spectrum.

According to the fundamental principles of statistical mechanics, the probability $w$ for a quantity to have a value in the interval from $r$ to $r+dr$ is proportional to $e^{\Delta S_t(r)}$, where $\Delta S_t(r)$ is the change in the entropy during the fluctuation (\citealt{LaLi}). The entropy change can be written as $\Delta S_t=-W_{\min}/T$, where $W_{\min}$ is the minimum work needed to carry out reversibly the given change in the thermodynamics quantities in a small part of the system. Thus $w\propto e^{-W_{\min}/T}$. For $W_{\min}$ we can adopt the expression
\begin{equation}
W_{\min}=\Delta E-T\Delta S+P\Delta V+\Delta R_{\min},
\end{equation}
where $\Delta E$, $\Delta S$, $\Delta V$ are the fluctuations in energy, entropy, and volume, respectively, while $\Delta R_{\min}$ is the minimal work necessary for reversible removal of mass $\Delta M$ for a distance $\delta r$ in the gravitational field of a mass $M$. Thus we have
\begin{equation}\label{fluct1}
w\propto \exp\left(-\frac{\Delta E-T\Delta S+P\Delta V}{T}-\frac{\Delta R_{\min}}{T}\right).
\end{equation}
Eq.~(\ref{fluct1}) can be applied to any fluctuation, small or large. However, in the following we consider only the case of small fluctuations. In this we can expand $\Delta E$ is series, obtaining
\begin{equation}
\Delta E=\frac{\partial E}{\partial S}\Delta S+\frac{\partial E}{\partial V}\Delta V +\frac{1}{2}\left[\frac{\partial ^2 E}{\partial S^2}\left(\Delta S\right)^2+2\frac{\partial ^2E}{\partial S \partial V}\Delta S \Delta V+\frac{\partial ^2E}{\partial V^2}\left(\Delta V\right)^2\right].
\end{equation}
By taking into account that $\frac{\partial E}{\partial S}=T$, $\frac{\partial E}{\partial V}=-P$, we obtain
\begin{eqnarray}
\Delta E-T\Delta S+P\Delta V&=&\frac{1}{2}\left[\frac{\partial ^2 E}{\partial S^2}\left(\Delta S\right)^2+2\frac{\partial ^2E}{\partial S \partial V}\Delta S \Delta V+\frac{\partial ^2E}{\partial V^2}\left(\Delta V\right)^2\right]=\nonumber\\
&&\frac{1}{2}\left(\Delta S\Delta T-\Delta P \Delta V\right),
\end{eqnarray}
giving
\begin{equation}
w\propto \exp\left(\frac{\Delta S\Delta T-\Delta P \Delta V}{2T}-\frac{\Delta R_{\min}}{T}\right).
\end{equation}

By taking $V$ and $T$ as independent variables, we obtain first
\begin{equation}
\Delta S=\left(\frac{\partial S}{\partial T}\right)_V\Delta T+\left(\frac{\partial S}{\partial V}\right)_T\Delta V=\frac{C_v}{T}\Delta T+\left(\frac{\partial P}{\partial T}\right)_V\Delta V,
\end{equation}
\begin{equation}
\Delta P=\left(\frac{\partial P}{\partial T}\right)_V\Delta T+\left(\frac{\partial P}{\partial V}\right)_T\Delta V,
\end{equation}
where $C_v$ is the specific heat at constant volume. Therefore for the fluctuation probability we obtain
\begin{equation}
w\propto \exp\left[-\frac{C_v}{2T^2}\left(\Delta T\right)^2+\frac{1}{2T}\left(\frac{\partial P}{\partial V}\right)_T\left(\Delta V\right)^2-\frac{\Delta R_{\min}}{T}\right].
\end{equation}

In order to estimate $R_{\min}$ we consider that it can be obtained as (\citealt{Roz2, Roz3})
\begin{equation}
\Delta R_{\min}\approx G\frac{M\Delta M}{r^2}\Delta r=\frac{4\pi }{3}G\rho r\Delta M \Delta r,
\end{equation}
where $M=4\pi \rho r^3/3$ is the mass at radius $r$, and $\Delta M=4\pi \rho r^2 \Delta r$ is the mass of a spherical shell of thickness $\Delta r$. Therefore the probability of thermal fluctuations in a self-gravitating system takes the form (\citealt{Roz2, Roz3})
\begin{equation}
w\propto \exp\left[-\frac{1}{2\delta _T^2}\left(\frac{\Delta T}{T}\right)^2-\frac{1}{2\delta _r^2}\left(\frac{\Delta r}{r}\right)^2\right],
\end{equation}
where $\delta _T^2=C_v^{-1}$, and
\begin{equation}
\delta _r^2=\left\{\frac{12\pi \rho }{kT}\left[\left(\frac{\partial P}{\partial V}\right)_Tr^3+\frac{8\pi G\rho }{9}r^5\right]\right\}^{-1},
\end{equation}
where $k$ is Boltzman's constant.
The root-sum-square relative density fluctuation (fluctuation spectrum) is given by (\citealt{Roz2})
\begin{equation}
\sqrt{\left\langle \left(\frac{\Delta \rho }{\rho }\right)^2\right \rangle}=3\sqrt{\left\langle \left(\frac{\Delta r }{r}\right)^2\right \rangle}=3\left\{\frac{12\pi \rho }{kT}\left[\left(\frac{\partial P}{\partial V}\right)_Tr^3+\frac{8\pi G\rho }{9}r^5\right]\right\}^{-1/2}.
\end{equation}
 The fluctuations of the density are related to the fluctuations of the radial coordinate by
\begin{equation}\label{13}
\frac{\Delta \rho }{\rho }=-3\frac{\Delta r}{r}.
\end{equation}
Since our main goal is to study the time variability of some astrophysical phenomena, we have to convert the radial variability into time variability. There are two possible time-scales for astrophysical phenomena, namely, the oscillation time scale and the free-fall time scale. For a homogeneous uniform self-gravitating system that behaves as a perfect gas throughout, the pulsation period $t _p $ is $t _p=2\pi/\sqrt{3\pi G\rho }$, while the free-fall time is $t _{ff}=\sqrt{4\pi/3G\rho }$. In both cases the time-scale is proportional to $1/\sqrt{\rho }$. Therefore, assuming such a general time - density dependence, we obtain
\begin{equation}
\frac{\Delta t}{t}=-\frac{1}{2}\frac{\Delta \rho }{\rho }.
\end{equation}
With the use of Eq.~(\ref{13}) we obtain
\begin{equation}
\frac{2}{3}\frac{\Delta t}{t}=\frac{\Delta r}{r},
\end{equation}
giving
\begin{equation}\label{time}
r\propto t^{2/3}.
\end{equation}
We can now discuss the limiting cases of the time-spectrum in a self-gravitating system. At spatial scales when the pressure gradients dominate, we have
\begin{equation}\label{eq1}
 \sqrt{\left\langle \left(\frac{\Delta \rho }{\rho }\right)^2\right \rangle}\approx 3\left\{\frac{12\pi \rho }{kT}\left[\left(\frac{\partial P}{\partial V}\right)_Tr^3\right]\right\}^{-1/2}\propto r^{-3/2}\propto t^{-1}.
 \end{equation}
 At large scales, when gravitational effects are important, we obtain
 \begin{equation}\label{eq2}
\sqrt{\left\langle \left(\frac{\Delta \rho }{\rho }\right)^2\right \rangle}\approx 3\left\{\frac{12\pi \rho }{kT}\left[\frac{8\pi G\rho }{9}r^5\right]\right\}^{-1/2}\propto r^{-5/2}\propto t^{-5/3}.
\end{equation}

However, the above discussion is incomplete in the sense that in a typical fluctuating system we have two independent modes, one growing, and one decaying. A better estimation of the time dependence of the fluctuations can be obtained as follows (\citealt{Roz4}).  According to Chebyshev's inequality from the probability theory, the probability that a density fluctuation $\Delta \rho /\rho$ exceeds $\delta >0$ is given by (\citealt{Fel})
\begin{equation}
P\left(\left|\Delta \rho /\rho\right|\geq \delta \right)\leq \frac{C}{\delta ^2},
\end{equation}
where $C$ is a constant. Now let's assume that for the independent growing mode we have $\delta _1\sim t^{\alpha }$, while the decaying mode behaves as $\delta _2\sim t^{-\gamma}$, where $\alpha, \gamma >0$ are constants. Then from Chebyshev's inequality we find $P_1\left(\left|\Delta \rho /\rho\right|= \delta _1\right)\sim t^{-2\alpha }$, and $P_2\left(\left|\Delta \rho /\rho\right|= \delta _2\right)\sim t^{2\gamma }$, respectively. We now take into account the collective effects, and look for the probability for the existence of a growing density contrast if there are neighboring regions of space in which the density inhomogeneity decreases. From a mathematical point of view we need to find the conditional probability $P_{12}$, which, according to the rules of probability theory, is given by $P_1=P_{12}P_2$. Therefore
\begin{equation}
P\left(\left|\Delta \rho /\rho\right|= \delta \right)=P_{12}=\frac{P_1}{P_2}\sim t^{-2\alpha -2\gamma }.
\end{equation}
By using again Chebyshev's inequality we find
\begin{equation}\label{inter}
\frac{\Delta \rho }{\rho }\sim t^{\alpha +\gamma }.
\end{equation}

\section{Fractal dimension and predictability index}\label{sect2}

The fractal dimension can be defined by the self-similar power law scaling
function (\citealt{Man, Pred})
\begin{equation}\label{2}
y(x)=a_sx^D,D>0,
\end{equation}
where $y(x)$ denotes the number of self-similar objects in the sphere or circle of a radius $x$; $a_s$  and $D$ stand for the scaling factor and for the spatial fractal dimension, respectively. Physical systems evolve not only in space but also in time. Therefore many time dependent natural processes  can be successfully fitted by the
temporal counterpart of the fractal function given by Eq.~(\ref{2}),
\begin{equation}
y(t)=a_t t^{D_t}, t > 0,
\end{equation}
in which $y(t)$ characterizes the time-evolution of the system, $D_t$ is its temporal fractal dimension whereas $a_t$ is a scaling factor.

In the following we will obtain an expression for the calculation of the fractal dimension by starting from the definition of the  Haussdorf dimension (\citealt{Sevcik, Sevcik1}). The Haussdorf $D_H$ dimension of a set in a metric space is defined as
 \begin{equation}\label{2_1}
 D_H=-\lim _{\epsilon\rightarrow 0} \frac{\ln \left[N\left(\epsilon \right)\right]}{\ln \epsilon},
 \end{equation}
 where $N\left(\epsilon \right)$ is the number of open balls of a radius $\epsilon $ needed to cover the entire set. In a metric space,
given any point $P$, an open ball of center $P$ and radius $\epsilon $, is a set of all points $x$ for which ${\rm dist}\left(P,x\right) <\epsilon$. A line of length $L$ may be divided into $N(\epsilon) = L/\left(2 \epsilon\right)$ segments of length $2\epsilon$, and may be covered by $N$ open balls of radius $\epsilon $. Thus, Eq.~(\ref{2_1}) may be rewritten as
\begin{equation}\label{3}
D_H=\lim _{\epsilon\rightarrow 0}\left[\frac{-\ln (L)+\ln \left(2\epsilon \right)}{\ln \left(\epsilon \right)}\right]=\lim _{\epsilon\rightarrow 0}\left[1-\frac{\ln (L)-\ln \left(2\right)}{\ln \left(\epsilon \right)}\right]=\lim _{\epsilon\rightarrow 0}\left[1-\frac{\ln (L)}{\ln \left(\epsilon \right)}\right].
\end{equation}

Waveforms are planar curves in a space with coordinates usually having different units. Since the
topology of a metric space does not change under linear transformation, it is convenient  to linearly
transform a waveform into another in a normalized space, where all axes are equal. \cite{Sevcik} proposed to use two linear transformations that map the original waveform into another, embedded in an equivalent
metric space. The first transformation, normalizes every point in the abscissa as $x_i^{*}=x_i/x_{\max}$, where  $x_i$ are the original values of the abscissa, and $x_{\max }$ is the maximum $x_i$. The second transformation
normalizes the ordinate as $y_i^{*}=\left(y_i-y_{\max}\right)/\left(y_{\max}-y_{\min}\right)$, where $y_i$ are the original values of the ordinate, and $y_{\min }$ and $y_{\max }$ are the minimum and maximum
$y_i$, respectively. These two linear transformations map the $N$ points of the waveform into another that
belongs to a unit square. This unit square may be visualized as covered by a grid of $N \times N$ cells, $N$ of
them containing one point of the transformed waveform. Calculating $L$ of the transformed waveform
and taking $\epsilon  = 1/(2\times N')$, where $N'=N-1$,  Eq.~(\ref{3}) becomes (\citealt{Sevcik})
\begin{equation}\label{4}
D_H\approx D=1+\frac{\ln (L)}{\ln \left(2\times N'\right)}.
\end{equation}
The approximation for the calculation of $D_H$ improves as $N'\rightarrow \infty$.

If the fractal dimension $D$ for the time series is 1.5, there is no correlation between amplitude changes corresponding to two successive time intervals. Therefore, no trend in amplitude can be discerned from the time series, and hence the process is unpredictable. However, as the fractal dimension decreases to 1, the process becomes more and more predictable as it exhibits "persistence". That is, the future trend is more and more likely to follow an established trend.  As the fractal dimension increases from 1.5 to 2, the process exhibits "anti-persistence". That is, a decrease in the amplitude of the process is more likely to lead to an increase in the future. Hence, the predictability again increases (\citealt{Pred}).

Based on the fractal dimension we can introduce the predictability index $P$ of a given physical system, defined as (\citealt{Pred1}),
\begin{equation}
PI=2\left|D-1.5\right|,
\end{equation}
where $\left|\;\;\right|$ denotes the absolute value of the number $D$. We use absolute values since predictability increases in both the following cases - when the fractal dimension becomes less than 1.5 and when it becomes greater than 1.5. In the former case, we have correlation (persistence) behavior and in the latter case, anti-correlation (anti-persistence) behavior. However, in either case, the process becomes more predictable. Thus, use of absolute values ensures that a process with $D=1.3$ has the same predictability index as a process with $D= 1.7$.

\section{Fractal dimension, predictability index and power spectrum index of IDV light curves}\label{sect3}

\begin{figure}[tbp]
\includegraphics[width=2.55in, height=1.80in]{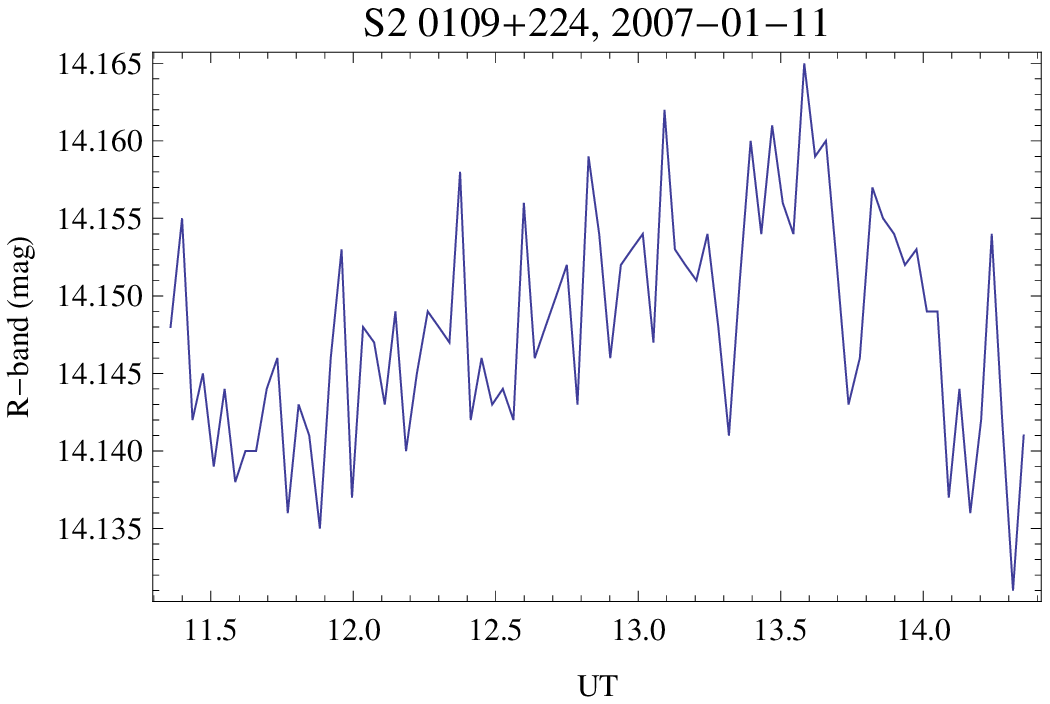}
\includegraphics[width=2.55in, height=1.80in]{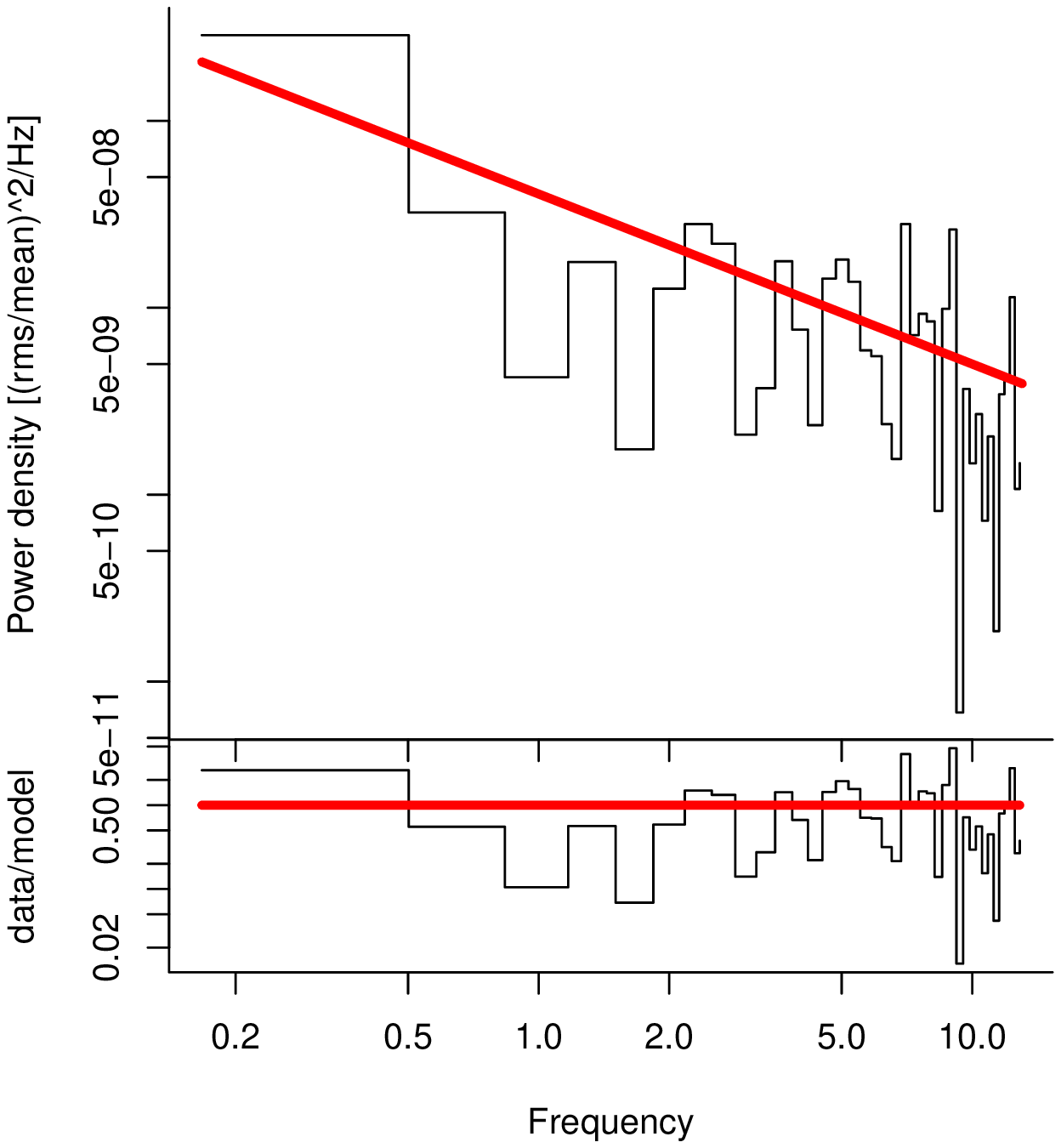}\\
\includegraphics[width=2.55in, height=1.80in]{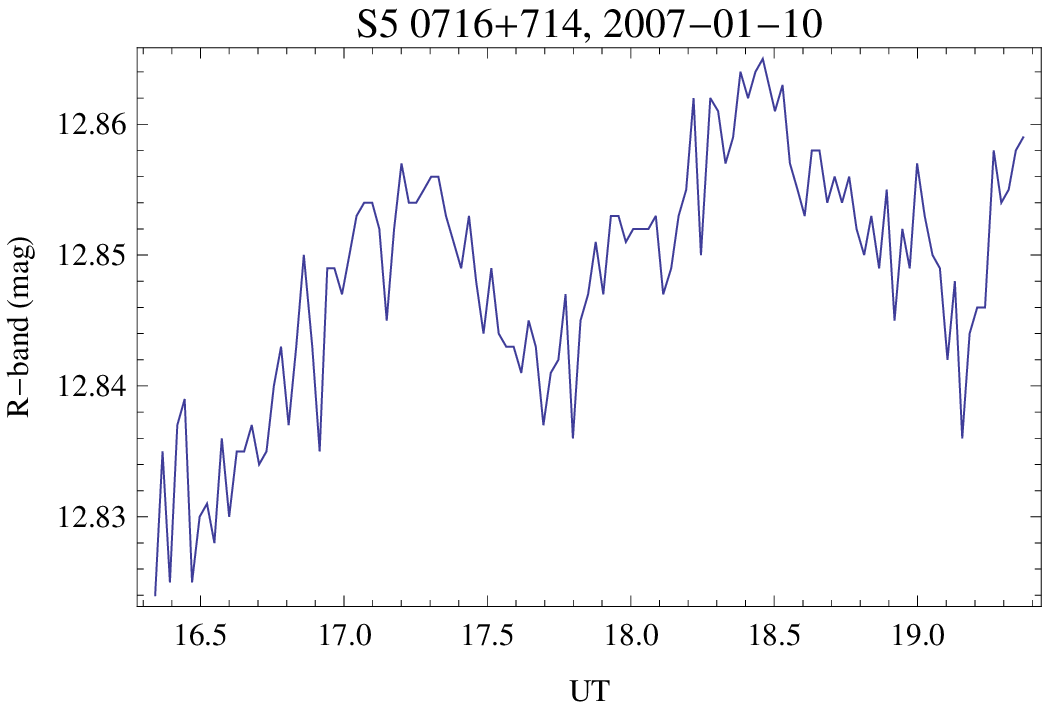}
\includegraphics[width=2.55in, height=1.80in]{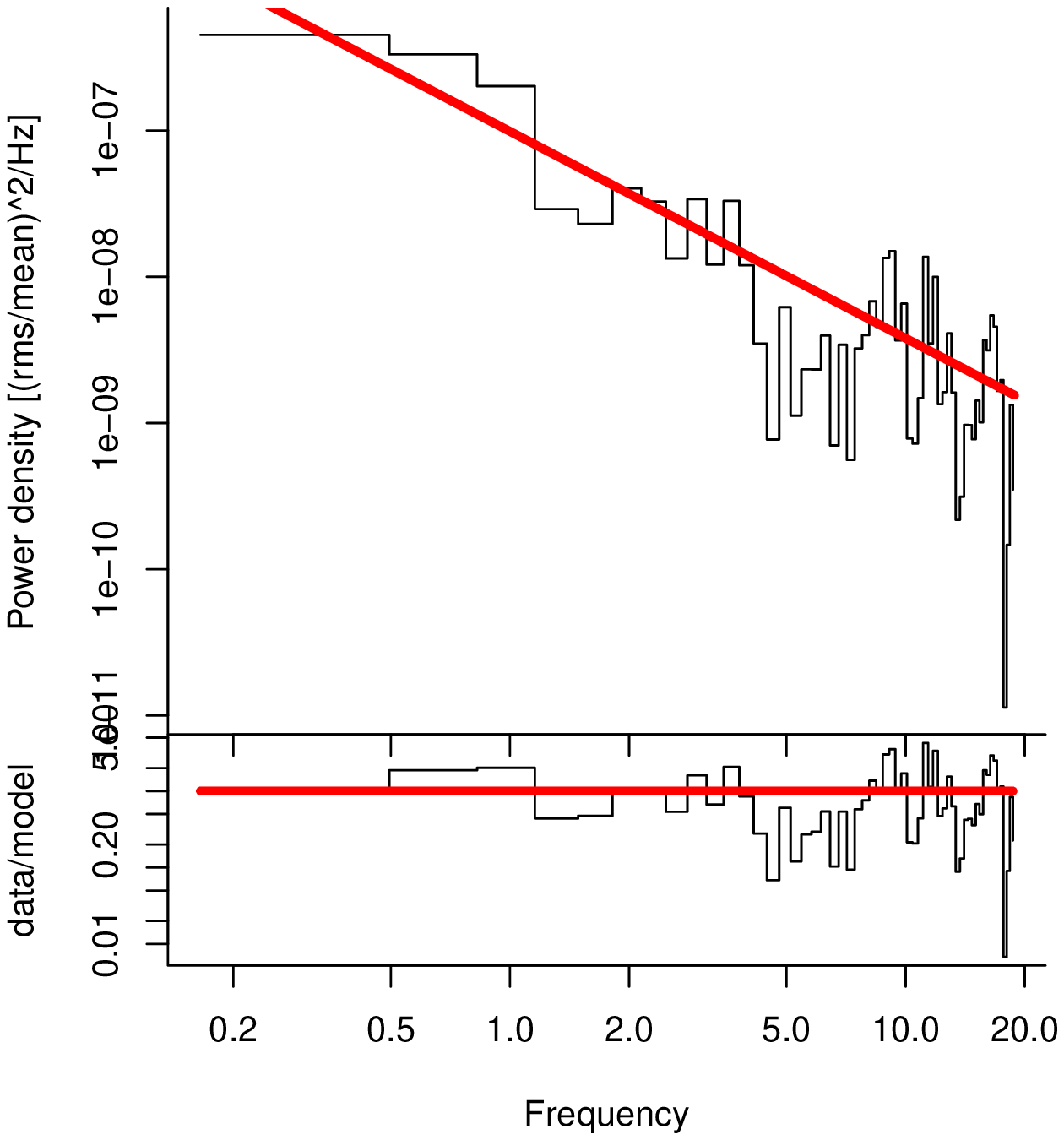}\\
\includegraphics[width=2.55in, height=1.80in]{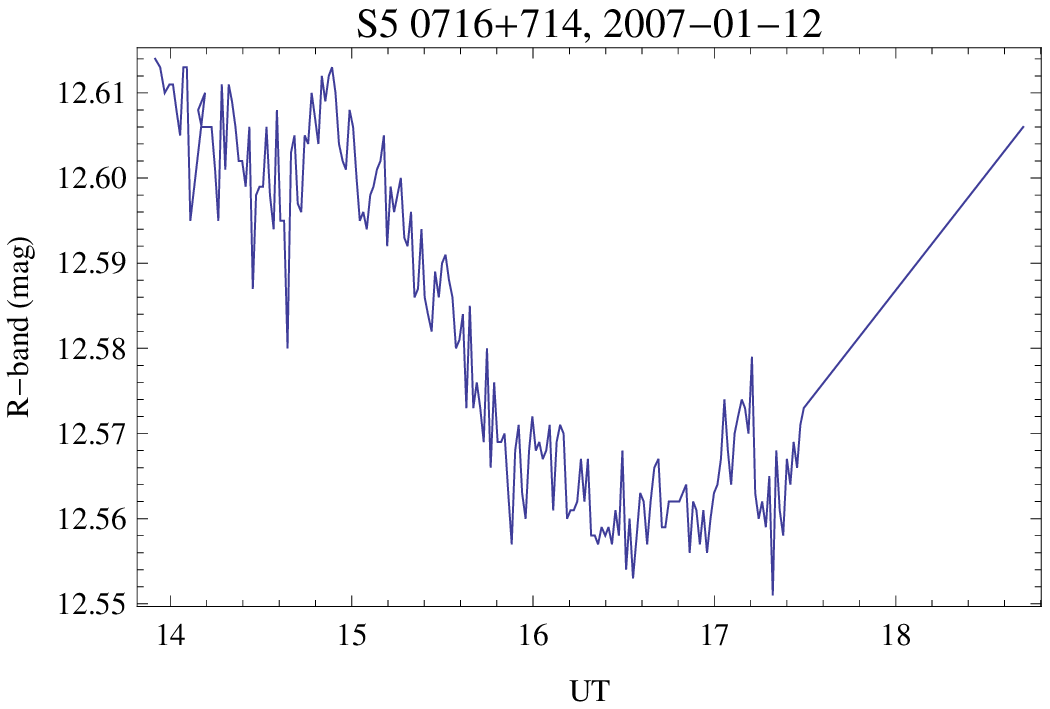}
\includegraphics[width=2.55in, height=1.80in]{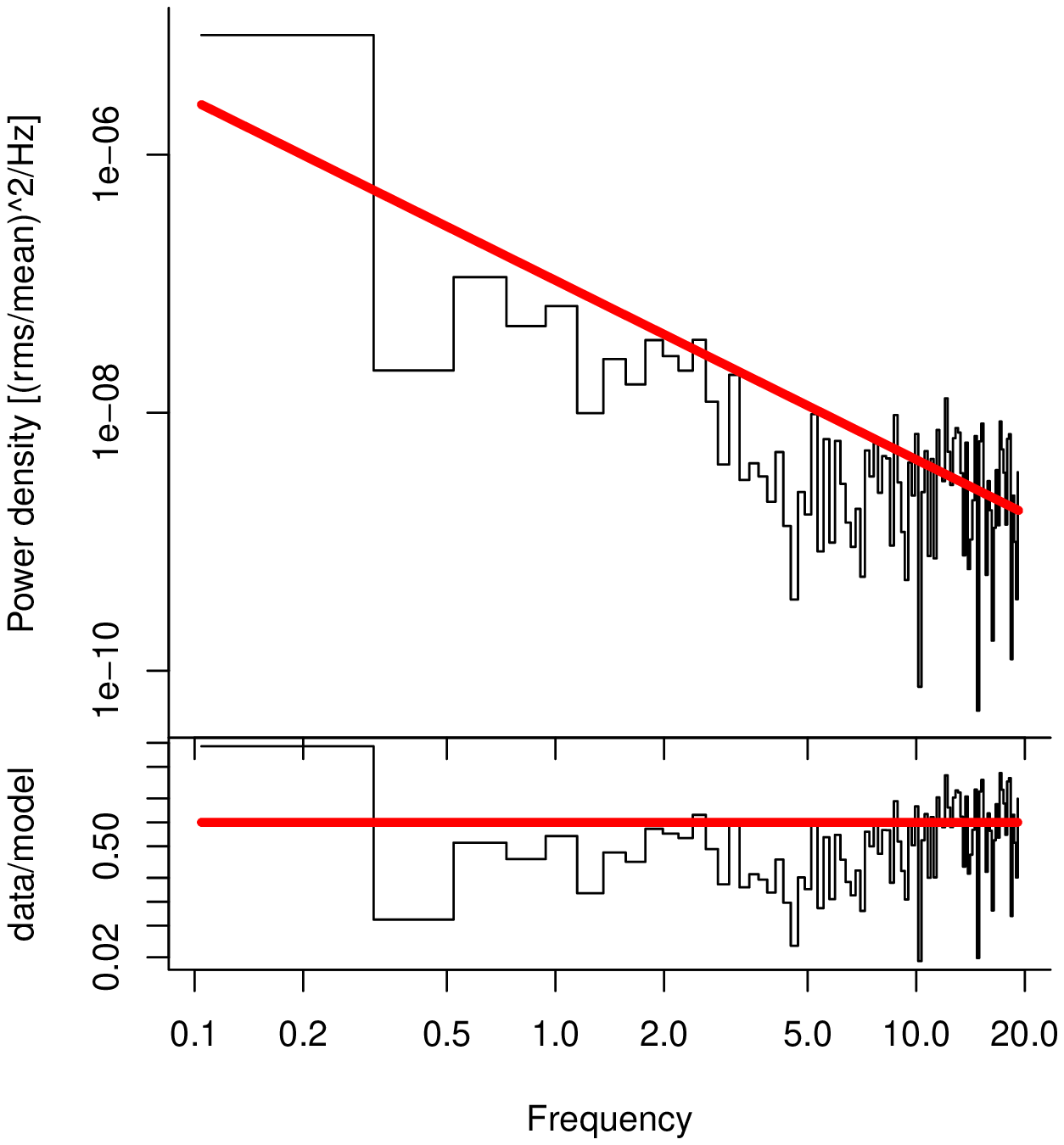}\\
\includegraphics[width=2.55in, height=1.80in]{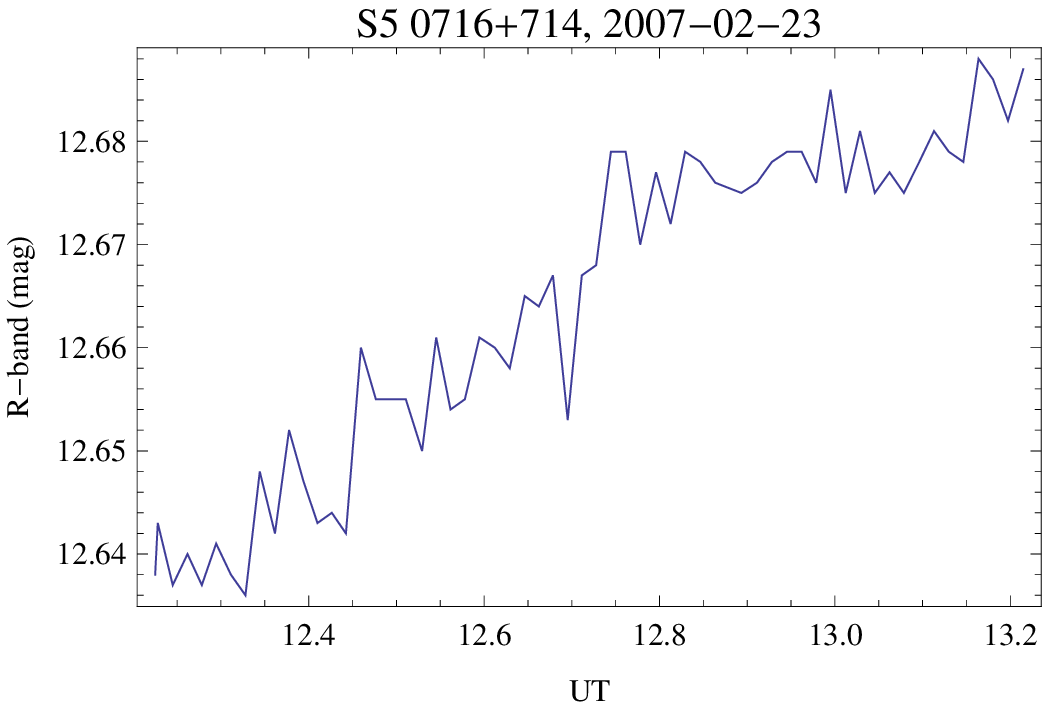}
\includegraphics[width=2.55in, height=1.80in]{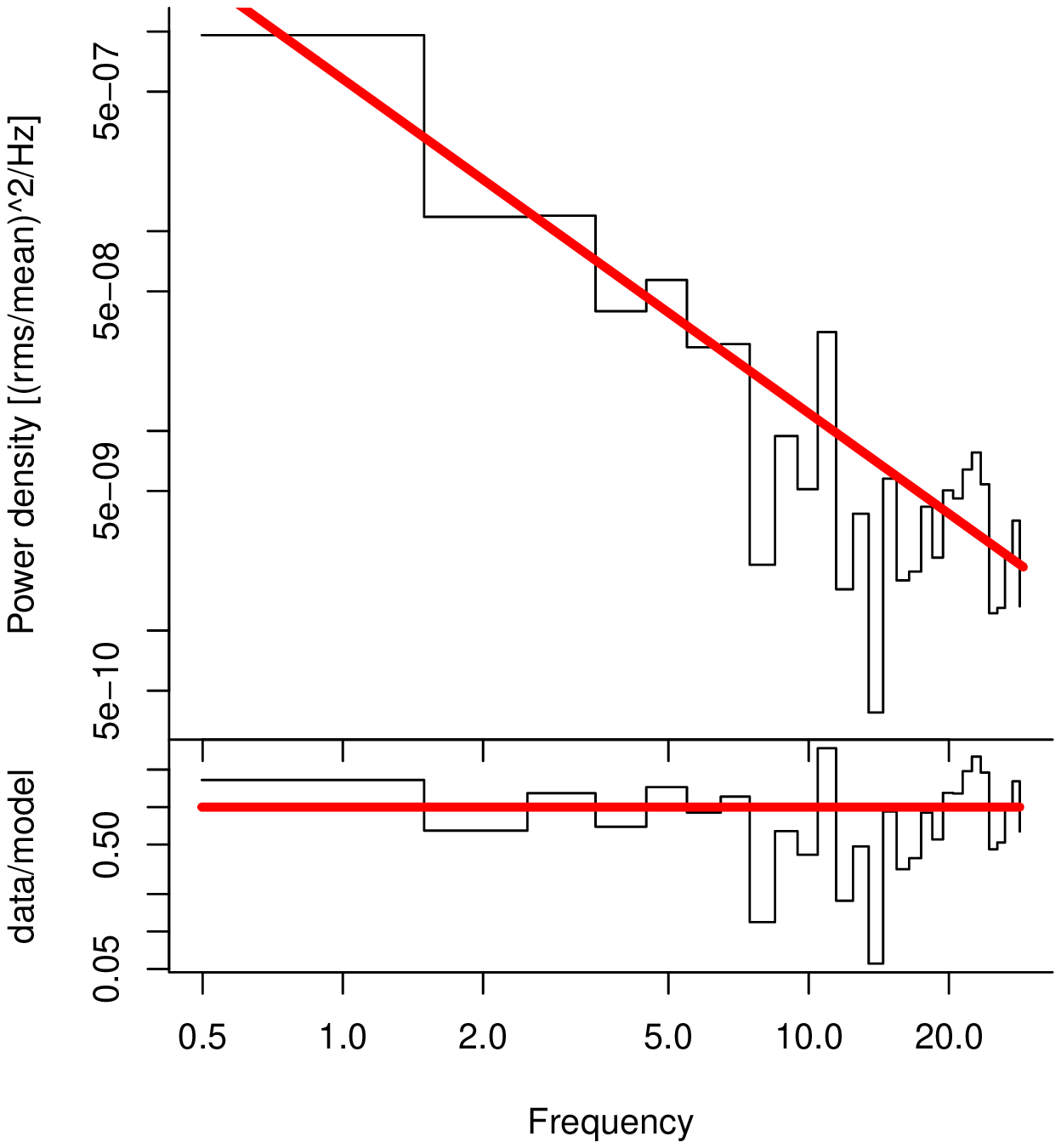}\\
\caption{R-band variability and power spectrum of the sample of blazars observed in \cite{Gu}.}
\label{fig1}
\end{figure}

\begin{figure}[tbp]
\includegraphics[width=2.55in, height=1.80in]{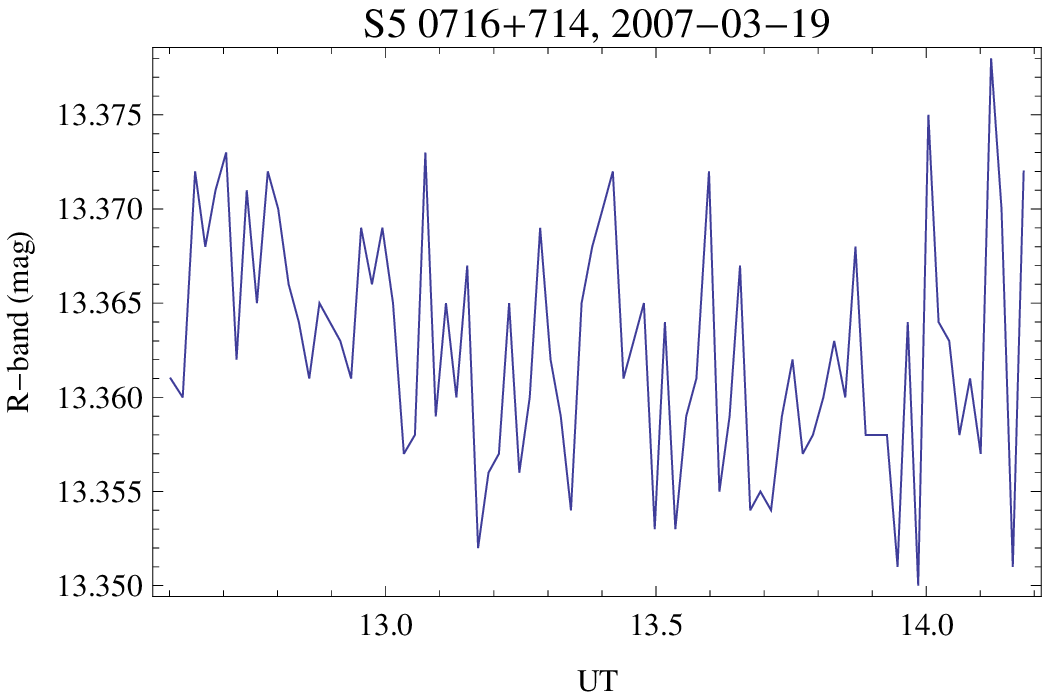}
\includegraphics[width=2.55in, height=1.80in]{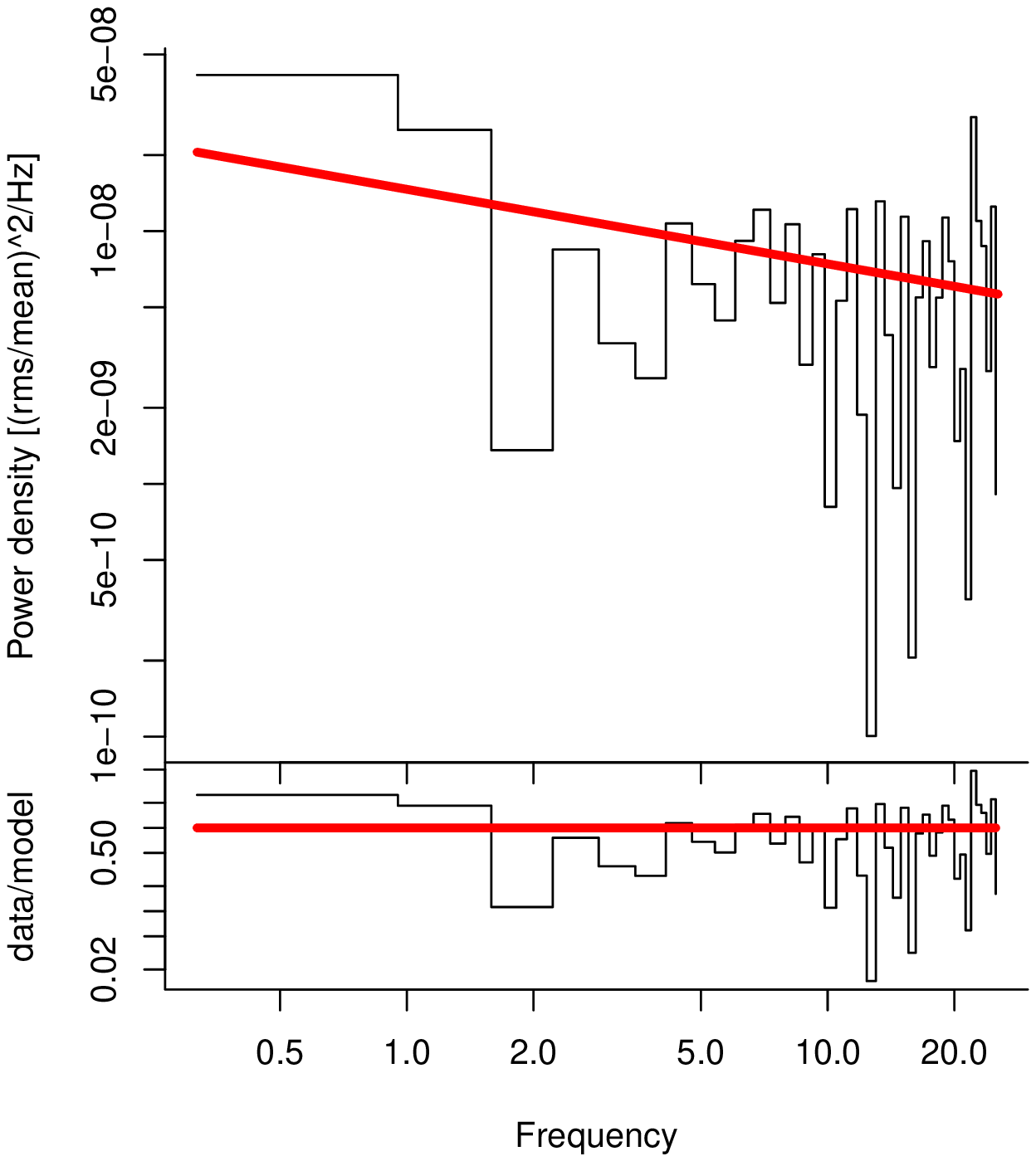}\\
\includegraphics[width=2.55in, height=1.80in]{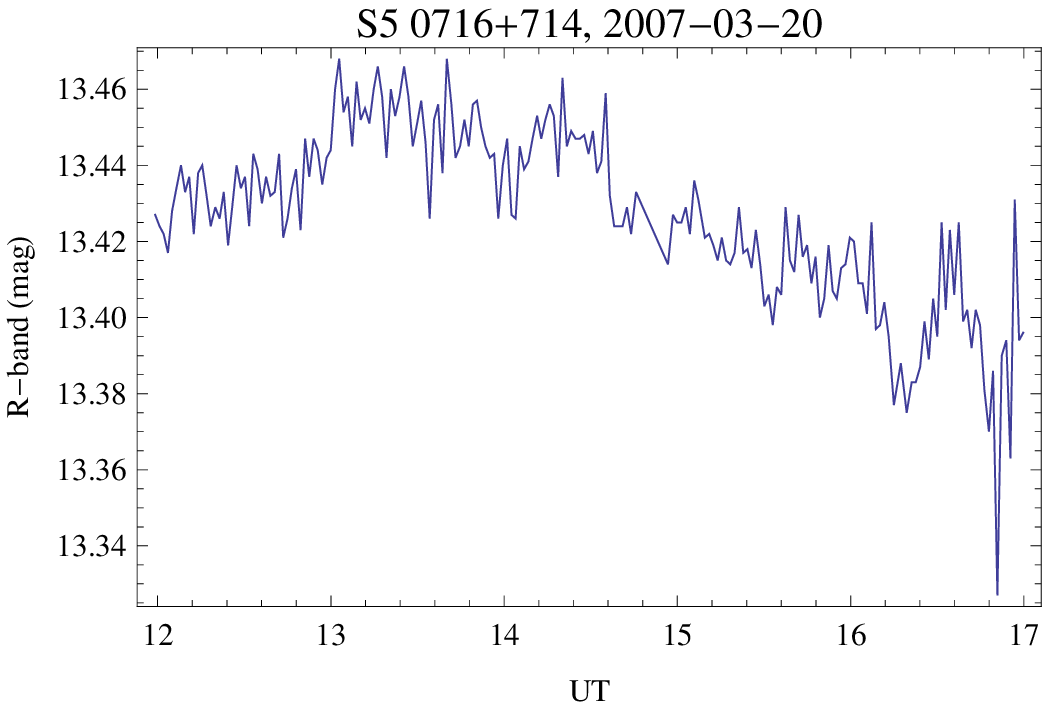}
\includegraphics[width=2.55in, height=1.80in]{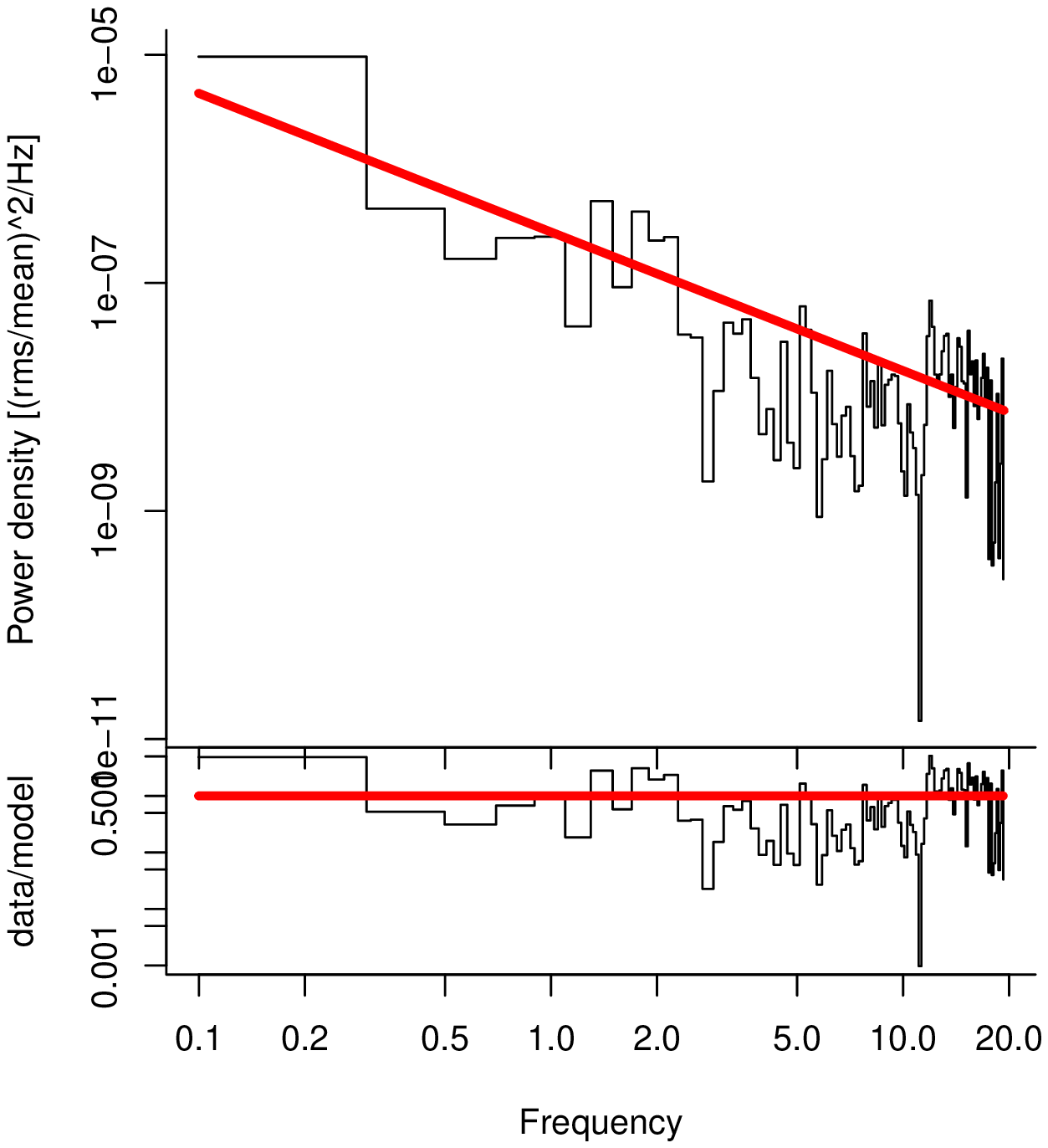}\\
\includegraphics[width=2.55in, height=1.80in]{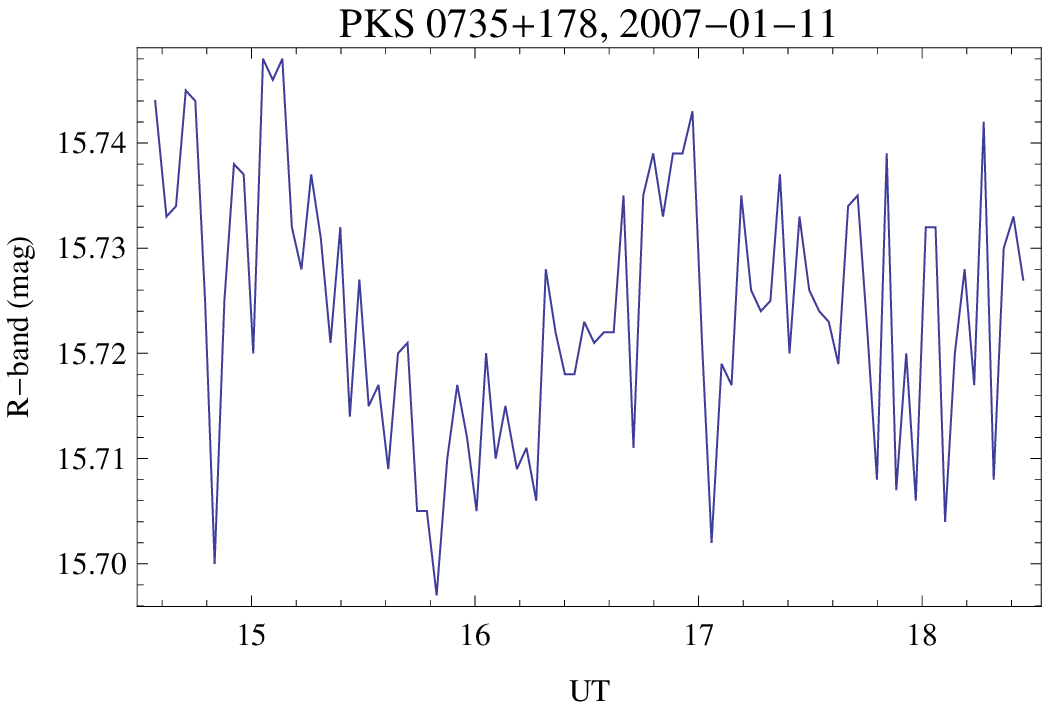}
\includegraphics[width=2.55in, height=1.80in]{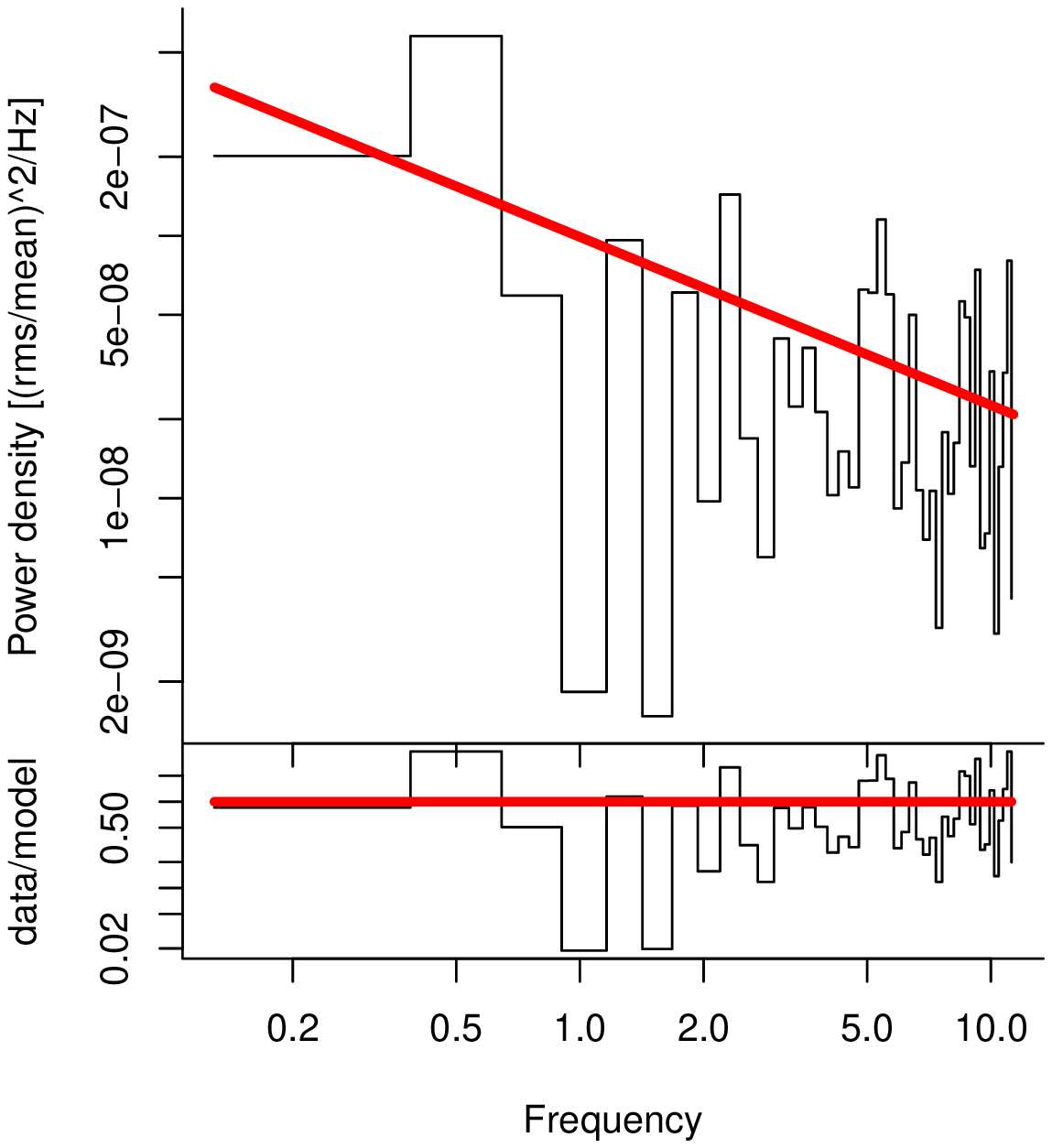}\\
\includegraphics[width=2.55in, height=1.80in]{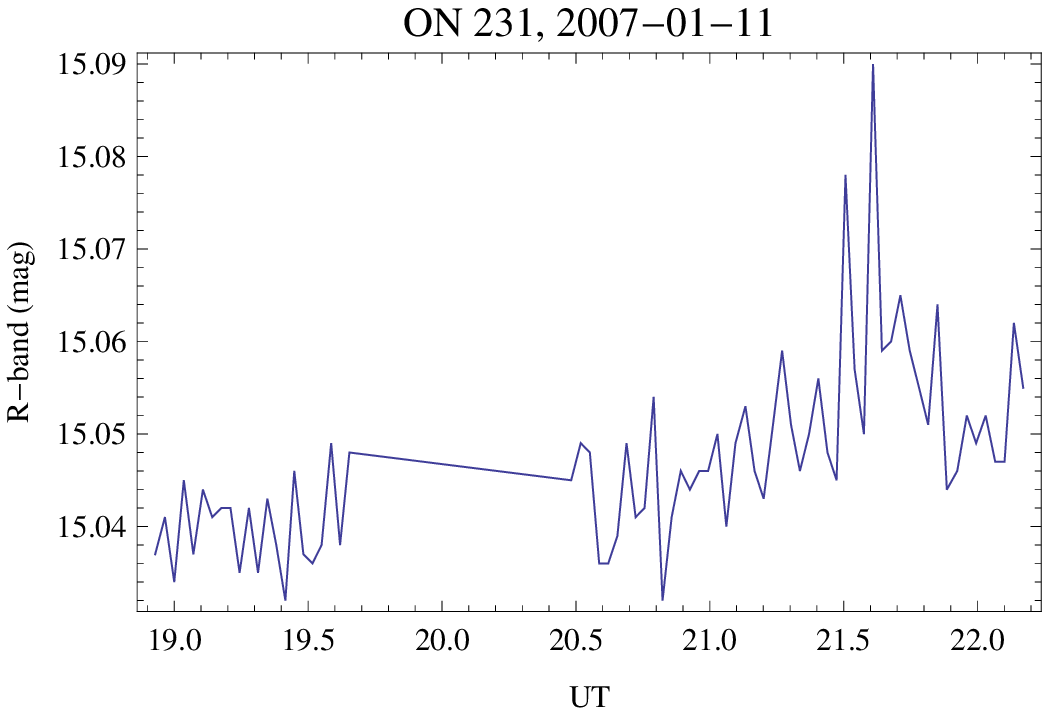}
\includegraphics[width=2.55in, height=1.80in]{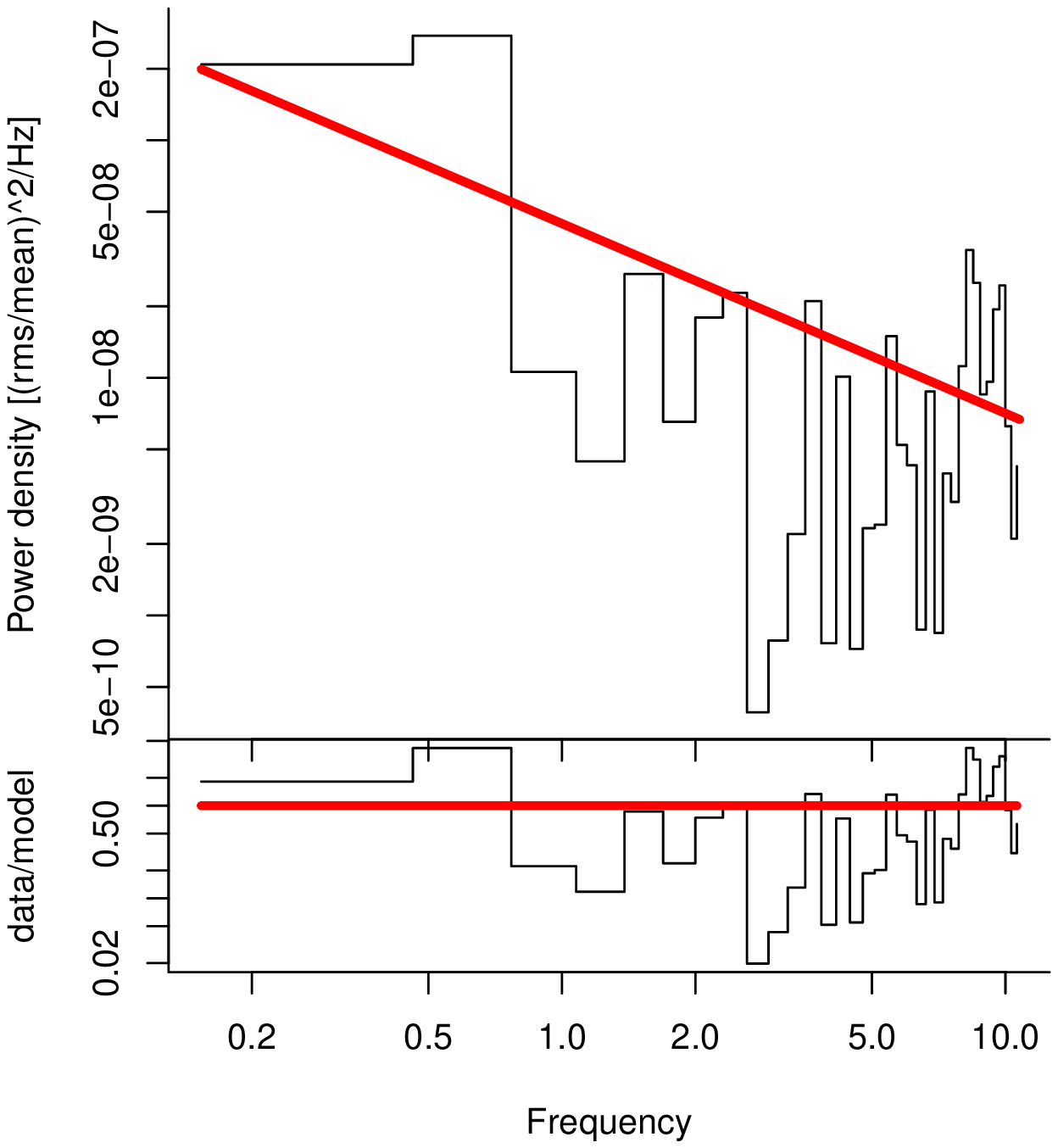}\\
\includegraphics[width=2.55in, height=1.80in]{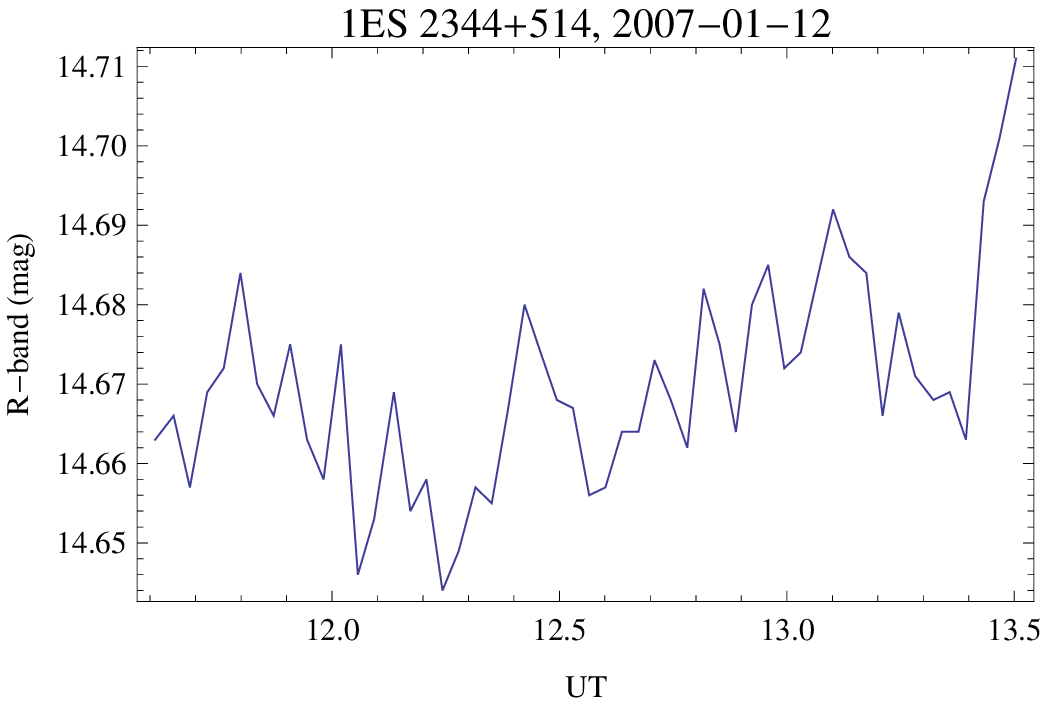}
\includegraphics[width=2.55in, height=1.80in]{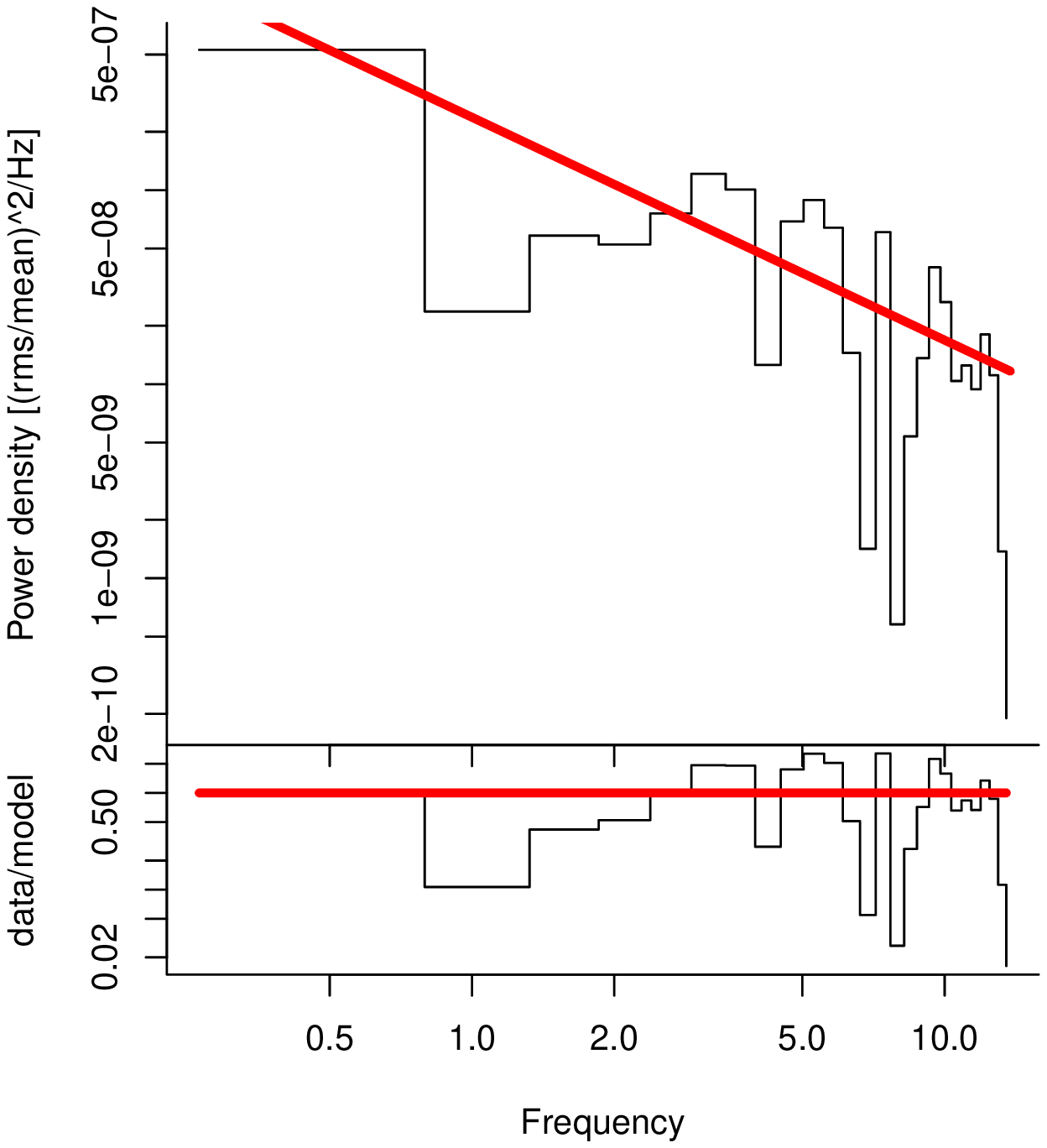}
\caption{R-band variability and power spectrum of the sample of blazars observed in \cite{Gu}.}
\label{fig2}
\end{figure}

In order to analyze the fractal properties of IDV light curves, we consider a sample of several blazars, which  have been observed in a simultaneous multi-wavelength observing campaign in their outburst phase by \cite{Gu}. Optical photometric monitoring of nine  blazars was carried out in 13 observing nights during the observing run of 2006 October 27-2007 March 20, using the 1.02 m optical telescope equipped with CCD detector and BVRI Johnson broadband filters at Yunnan Astronomical Observatory, Kunming, China. The R-band observations of five blazars are represented in Figs.~\ref{fig1} and \ref{fig2}, respectively.

There are many ways to characterize different noise sources. One possibility, which we have also explored in the present paper, is to consider the spectral density, that is, the mean square fluctuation at any particular frequency $f$, and its variation with frequency. In many natural processes the power spectrum $P(f )$ is proportional to $1 / f^{\beta }$, where the power index $\beta \geq 0$. When $\beta =0$, the noise is referred to as white noise, when it is 2 it is referred to as a Brownian noise, and when it is 1 it is called as the  $1/f$ noise (\citealt{Hav,Den}).

In order to obtain the power spectral index $\beta $ we have used the periodogram method, as developed in \cite{V05} and \cite{V2010}, respectively. To fit the data we have used power law fitting for  all our bands. In fact, we can also use other options.  However, since the number of available data points in the optical and radio bands is limited, for the sake of uniqueness we can only use power law fitting method.  Although the X-ray data can be fitted by different functions, the optical data and the radio data cannot be fitted by other methods, such as the bending power method. Therefore, we fitted  the data in all bands by using a power law. With the use of the above procedures, we can obtain the power spectrum, the power spectrum index, and all the other useful parameter information, such as the $p$-value. The error values after fitting have also been obtained, and we present them in Table~\ref{table1} and \ref{table2}, respectively, as a percentage error.

The fractal dimensions, the predictability index $PI_{IDV}$ and the power index of the observed blazar optical time series are represented in Table~\ref{table1}.

\begin{table}[h]
\begin{tabular}{|c|c|c|c|c|c|}
\hline
Object & Date of observation& $D$ & $PI_{IDV}$ & $\beta $ & Fit error of $\beta $\\
\hline
S2 0109+224 & 2007-01-11 & 1.52591 & 0.051 & 0.909647 & 0.1723201\\
\hline
S5 0716+714 & 2007-01-11 & 1.44996 & 0.100 & 1.41473 & 0.1404088 \\
\hline
S5 0716+714 & 2007-01-12 & 1.45699 & 0.086 & 1.389167 & 0.07811043\\
\hline
S5 0716+714 & 2007-02-23 & 1.35088 & 0.298 & 1.672849 & 0.2028525 \\
\hline
S5 0716+714 & 2007-03-19 & 1.57989 & 0.159 & 0.2951006 & 0.1540101  \\
\hline
S5 0716+714 & 2007-03-20 & 1.44484 & 0.110 & 1.21574 & 0.08695967 \\
\hline
PKS 0735+178 & 2007-01-11 & 1.56557 & 0.131 & 0.6411014 & 0.1548084 \\
\hline
ON 231 & 2007-01-11 & 1.45285 & 0.094 & 0.7989009 & 0.1536858\\
\hline
1ES 2344+514 & 2007-01-12 & 1.42142 & 0.157 & 1.148679 & 0.2762177\\
\hline
\end{tabular}
\caption{Fractal dimensions, predictability indices,  and power indices of a sample of R-band light curves (\citealt{Gu}).}
\label{table1}
\end{table}

As a second example of the determination of the fractal dimension we consider the short time-scale radio variations of the compact extragalactic radio source J 1128+5925 (\citealt{Gab}). The flux density variability of J 1128+5925 was monitored with dense time sampling between 2.7 and 10.45 GHz. For our analysis we use the data observed
with the 100 m Effelsberg radio telescope of the Max-Planck-Institut fur Radioastronomie at 2.70 GHz, 4.85 GHz,
and 10.45 GHz. Ten observing sessions that lasted several days during the period between 2004 - 2006 were performed.

Apart from the optical and radio data, we also try to use short time-scale of IDV data from XMM database for calculating the fractal dimension. We reduced the data and followed the result from the published targets as shown in \cite{Gau}. From the XMM database we have reduced 7 blazars.

We used {\it XMM-Newton} Science Analysis System (SAS) version 11.0 to reduce the EPIC-pn data taken from the archive. We followed the same procedure outlined in Gaur et al. (2010) to extract the lightcurves. In brief, we excluded all the high background due to flaring, and restricted all the events to the 0.3--10 keV energy band. We used 45 arcsec radius as the extraction region for both the source and the background (from source-free region). In addition, the background subtracted light curves were corrected for vignetting, bad pixels, Point Spread Function (PSF) variation, quantum efficiency, and dead time. All the light curves were binned with 100-second timing resolution.

\begin{figure}[tbp]
\includegraphics[width=2.55in, height=1.80in]{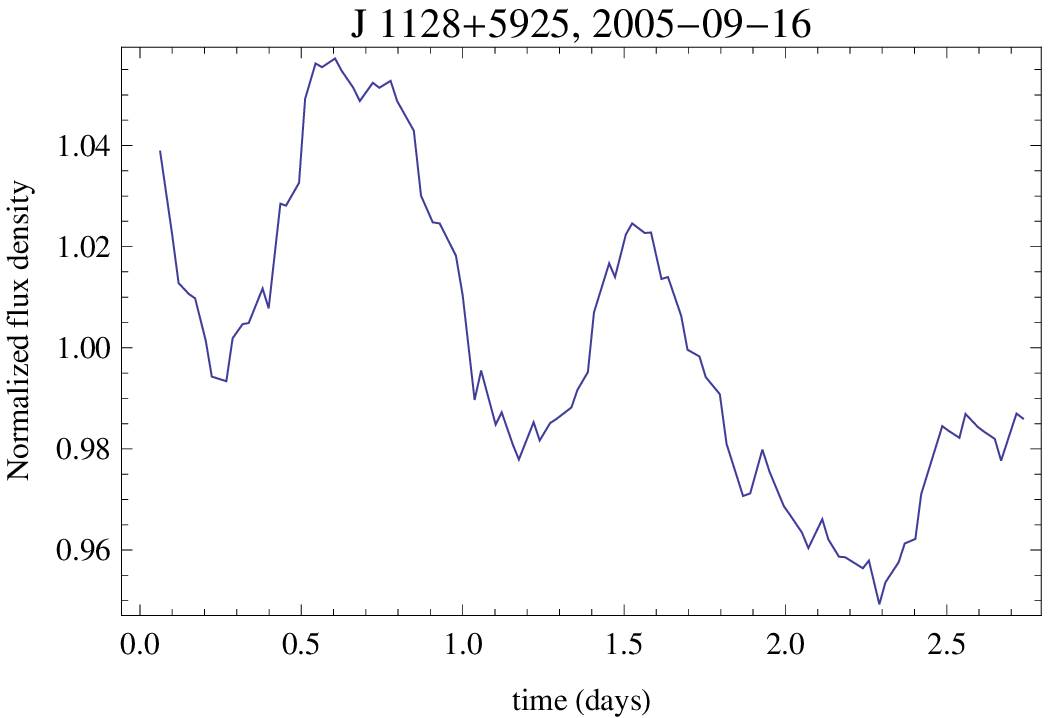}
\includegraphics[width=2.55in, height=1.80in]{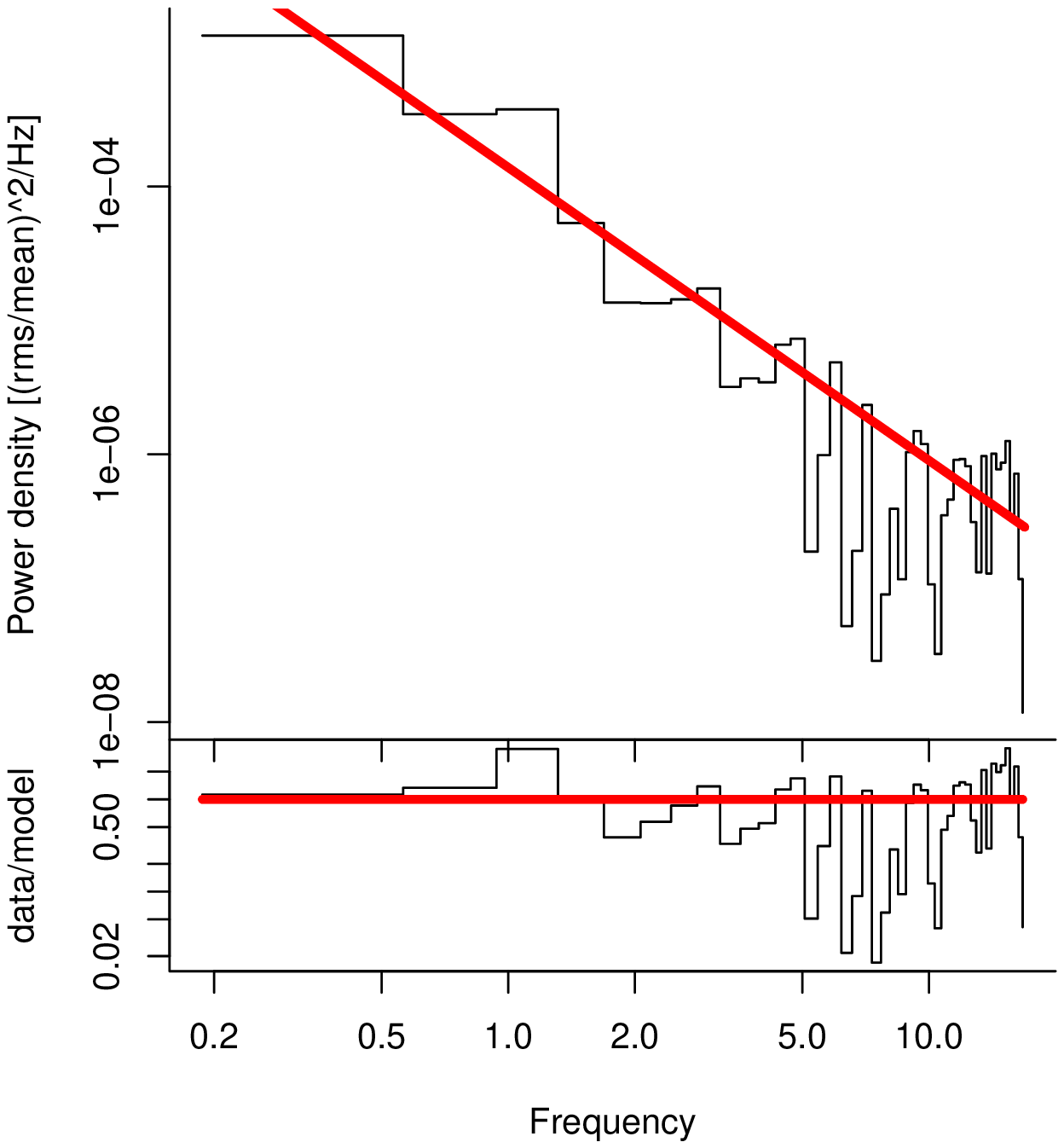}\\
\includegraphics[width=2.55in, height=1.80in]{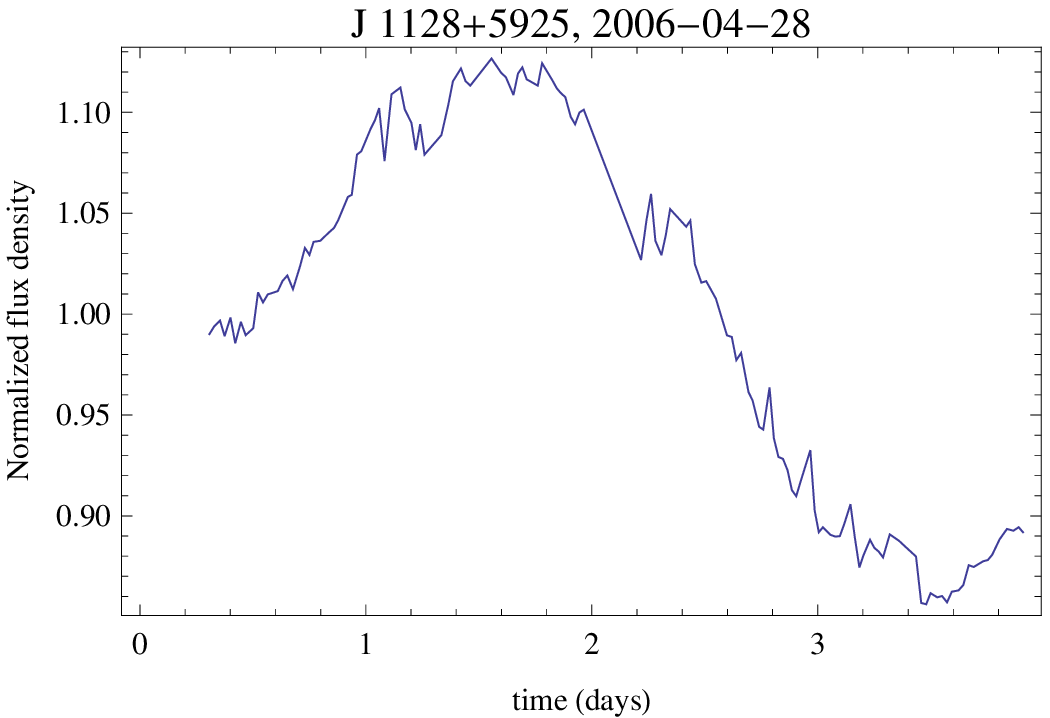}
\includegraphics[width=2.55in, height=1.80in]{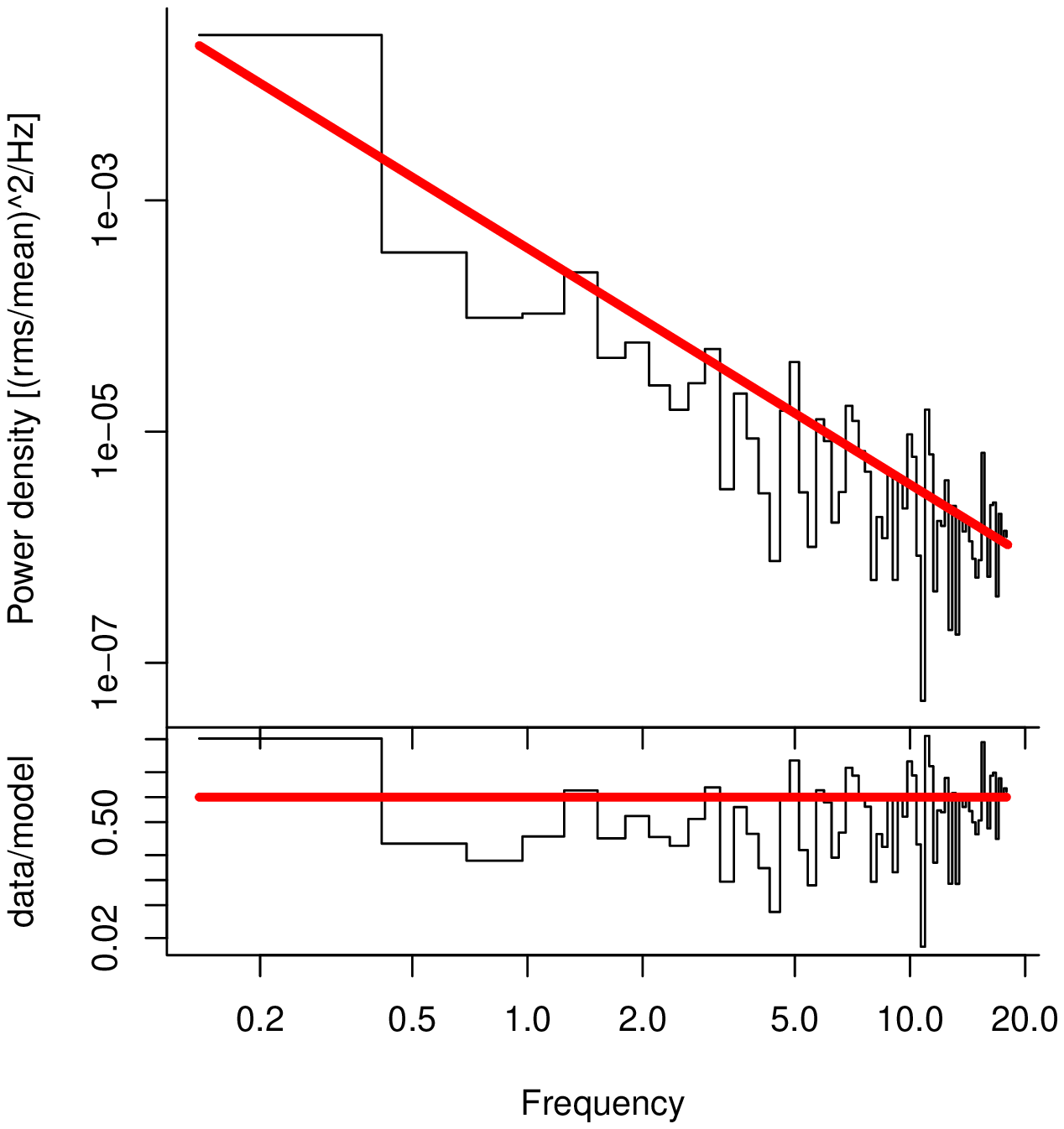}\\
\includegraphics[width=2.55in, height=1.80in]{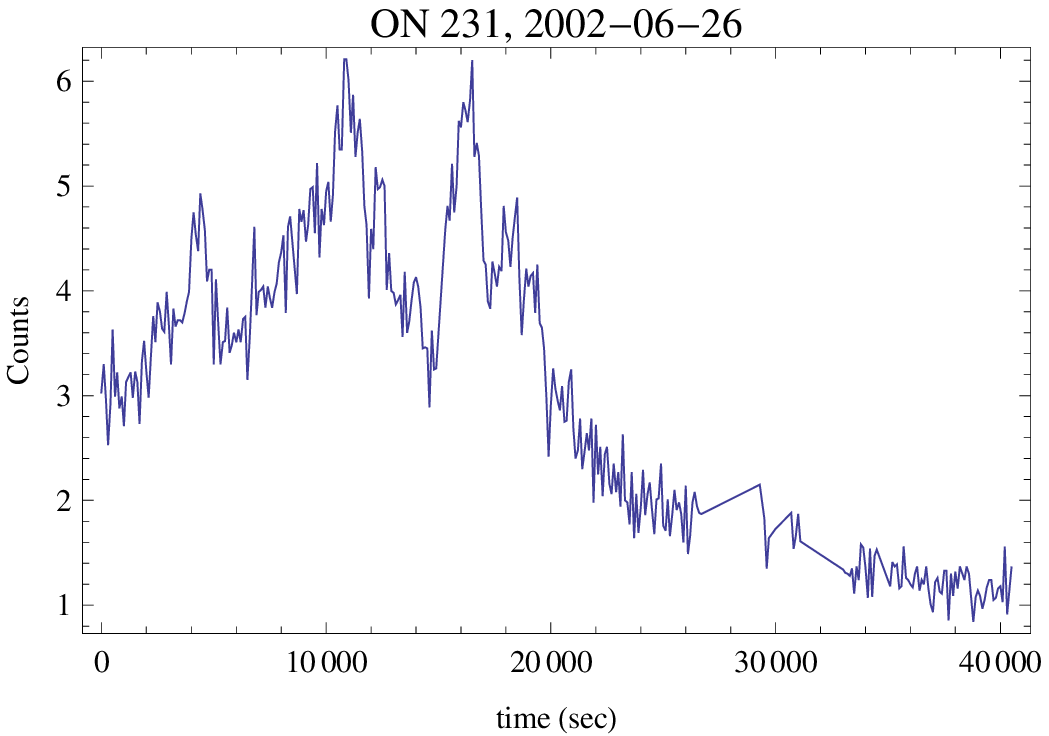}
\includegraphics[width=2.55in, height=1.80in]{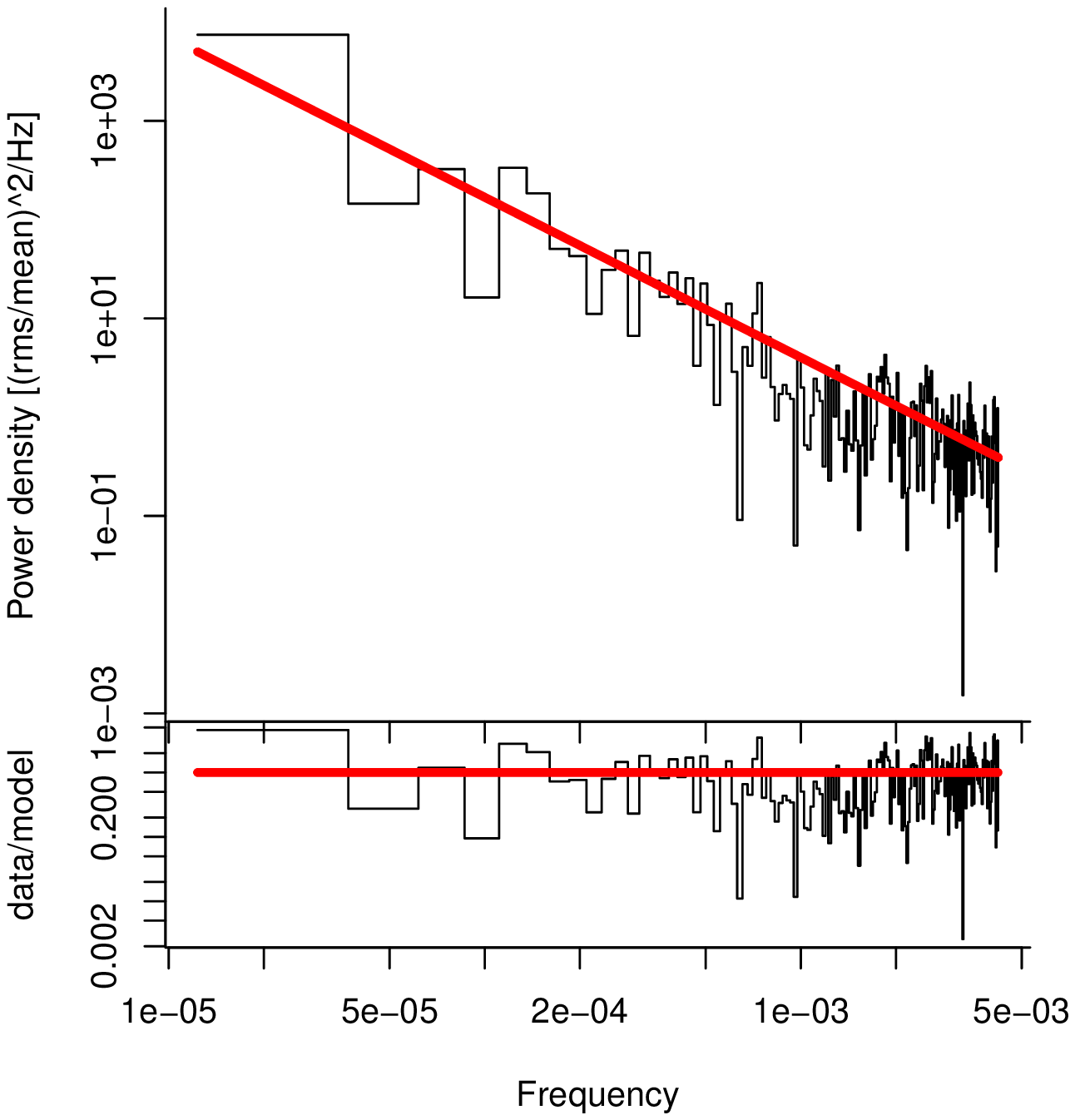}\\
\includegraphics[width=2.55in, height=1.80in]{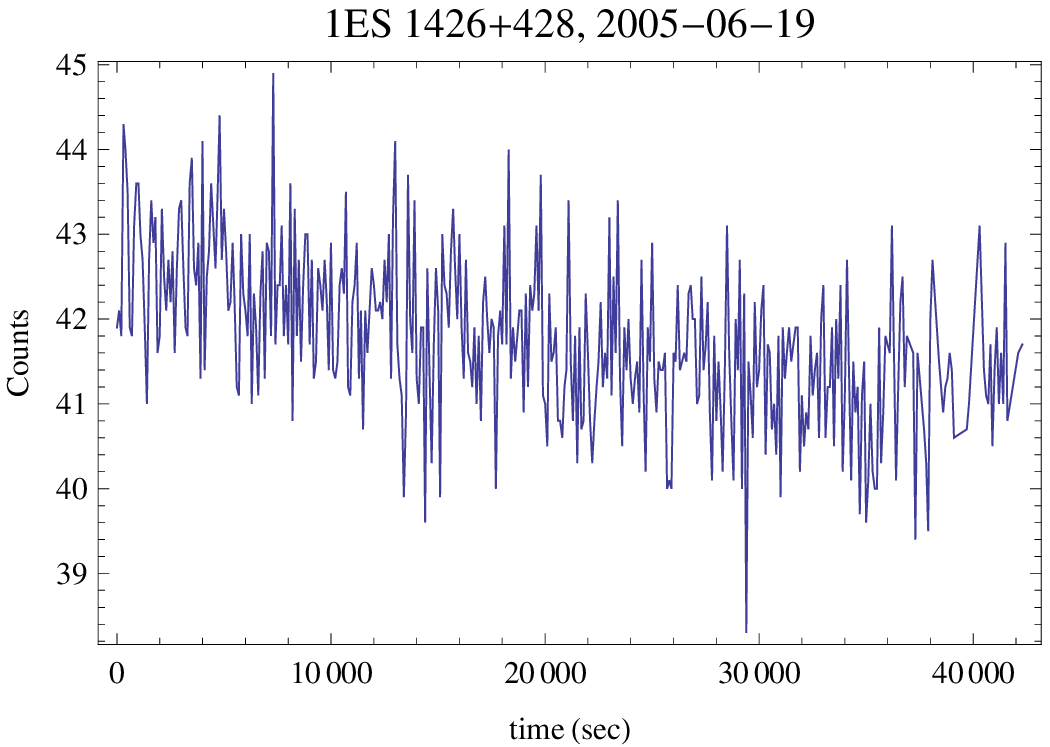}
\includegraphics[width=2.55in, height=1.80in]{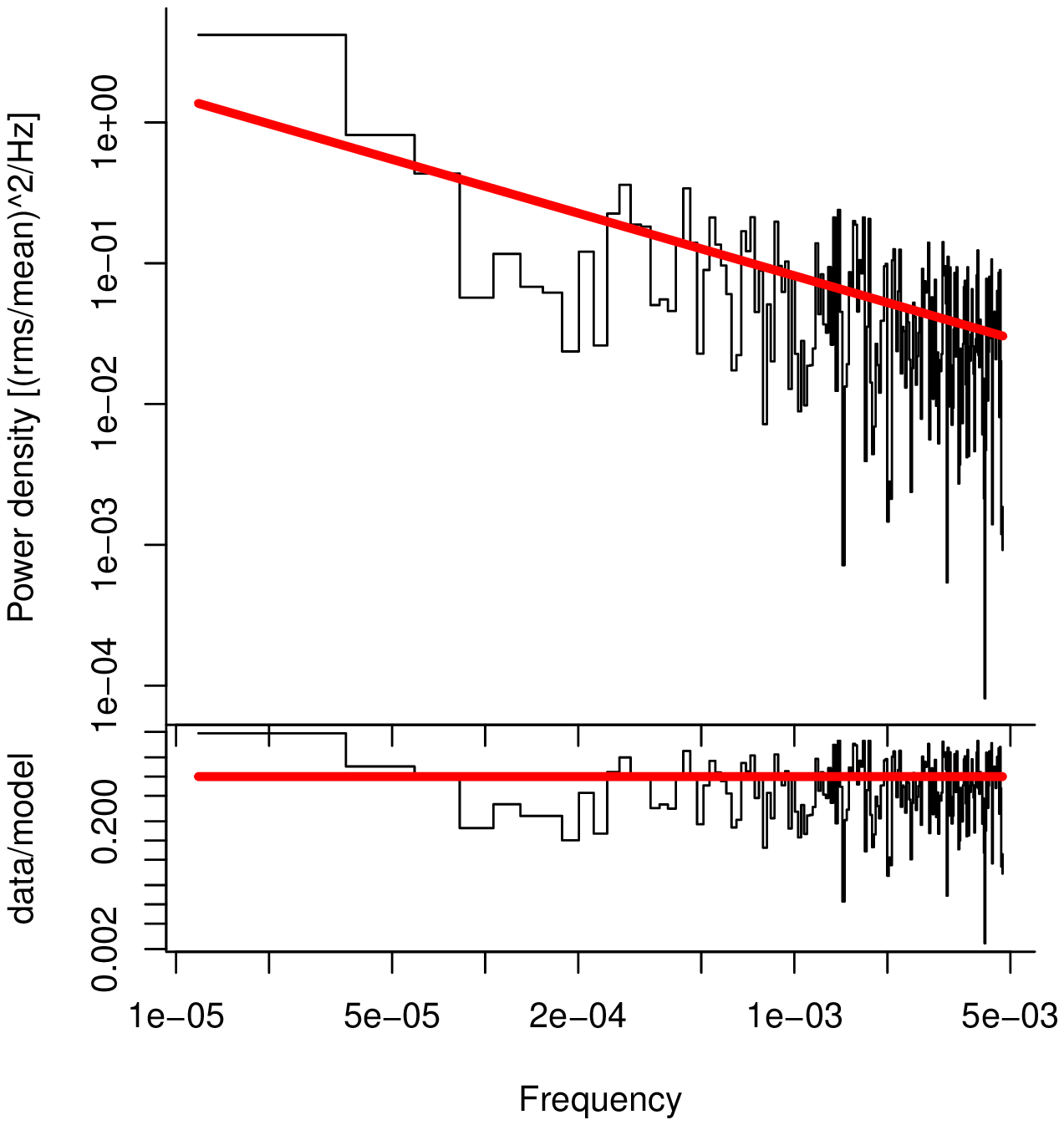}\\
\caption{Radio-band variability and power spectrum of the compact extragalactic radio source J 1128+5925 (\citealt{Gab}) and X-ray variability and power-spectrum of the sources  ON 231 and 1ES 2344+514, respectively.}
\label{fig3}
\end{figure}

The radio and X-ray data, together with the corresponding power spectra are shown in Fig.~\ref{fig3}, while the fractal dimension, predictability indices and the power spectral indices are presented in  Table~2.

\begin{table}[h]
\begin{tabular}{|c|c|c|c|c|c|}
\hline
Object & Date of observation& $D$ & $PI_{IDV}$ & $\beta $ & Fit error of $\beta $\\
\hline
J 1128+5925 & 2005-09-16 & 1.28642 & 0.42716 & 2.188192 & 0.1472077\\
\hline
J 1128+5925 & 2006-04-28 & 1.25587 & 0.48826 & 2.042647 & 0.1100852\\
\hline
ON 231 & 2002-06-26 & 1.43877 & 0.122 & 1.621563 & 0.06751574\\
\hline
1ES 2344+514 & 2005-06-19 & 1.60232 & 0.204 & 0.63486 & 0.06634579\\
\hline
PKS2155-304 & 2002-05-24 & 1.44012 &  0.11976 & 1.055890 & 0.0554563\\
\hline
PKS2155-304 & 2002-05-24 & 1.56946 & 0.13892 & 0.6418146 & 0.06857538\\
\hline
PKS2155-304 & 2002-05-24 & 1.33378 & 0.33244  & 1.615948 & 0.06764645\\
\hline
PKS2155-304 & 2000-05-31 & 1.57255 & 0.1451 & 1.023354 & 0.04338111\\
\hline
PKS2155-304 & 2000-11-21 & 1.51078 & 0.02156 & 1.171058 & 0.04761672\\
\hline
\end{tabular}
\caption{Fractal dimensions, predictability indices,  and power spectral indices of the compact extragalactic radio source J 1128+5925 (\citealt{Gab}) and of the X-ray sources  ON 231, 1ES 2344+514 and PKS 2155-304 respectively (\citealt{Gau}).}
\label{table2}
\end{table}

There is a relation between the fractal dimension $D$ of the Brownian noise and the power index $\beta $, namely (\citealt{Hav,Den})
\begin{equation}
D=\frac{5-\beta }{2}.
\end{equation}
If the fractal dimension of the considered IDV sources is around $D=1.5$, then the power spectral index of the spectra is around $\beta =2$. We have studied the correlation between the fractal dimension $D$ and the power spectral index of the optical and X-ray IDV sources - we did not include in our study the radio data, due to their small number. There is a very good correlation between $D$ and $\beta $, and for this 15 sources the relation between the fractal dimension and the power index can be given as
\begin{equation}\label{eqc}
D=1.66229-0.16699\times \beta.
\end{equation}

Eq.~(\ref{eqc}) can also be written, with a very good approximation, as
\begin{equation}
D=\frac{10-\beta }{6}.
\end{equation}

The correlation statistics of the $D-\beta $ correlation is represented in Table~\ref{table3}. For the chosen set of data the correlation coefficient between these two parameters is 0.8428.

\begin{table}[h]
\begin{tabular}{|c|c|c|c|}
\hline
C. C. & ${\rm C.C.}^2$ & Standard Error & Total number of cases \\
\hline
0.8428 & 0.7104 & 0.0469 & 15 \\
\hline
\end{tabular}
\caption{Regression statistics for the correlation between fractal dimensions and power spectral indices of optical and X-ray sources. The correlation coefficient is denoted by C.C.}
\label{table3}
\end{table}

The statistical significance of the correlation coefficient is presented in Table~\ref{table4} by means of the ANOVA analysis. In Table~\ref{table3}, C.C represents the correlation coefficient. It represents the fitting accuracy. With the C.C equal to 1 and -1 the scatter diagram can be fitted by one single straight line.  ${\rm C.C} ^2$ represents the square of C.C. In Statistics, the number of degrees of freedom is the number of values in the final calculation of a statistic that are free to vary.
If the F-statistics computed in the ANOVA table is less than the F-table statistics, or
the p-value is greater than the alpha  level of significance, then there is no reason to reject the null hypothesis
that all the means are the same.
For our p-level value, we obtained 0.00008, and thus we conclude that the null hypothesis should not be rejected at the 1\% significance level, since the p-level value is much less than 0.01.

\begin{table}[h]
\begin{tabular}{|c|c|c|c|c|c|}
\hline
Name & d.f. & SS & MS & F & p-level \\
\hline
Regression & 	1 &0.07041 &0.07041 & 31.89023 & 0.00008 \\
\hline
Residual & 13 & 0.0287 & 0.00221 & $-$ & $-$  \\
\hline
Total & 14 & 0.09912 & $-$ & $-$ & $-$   \\
\hline
\end{tabular}
\caption{ANOVA Table for the correlation coefficient of the fractal dimensions and power spectral indices of optical and X-ray sources. d.f. represents the number of degrees of freedom, SS is the sum of squares, MS is the mean sum squares value, F represents the value of the F-statistics, and p-level represents the measure of the degree of the insignificance of the regression model.}
\label{table4}
\end{table}

\begin{figure}[tbp]
\includegraphics[width=3.55in, height=2.80in]{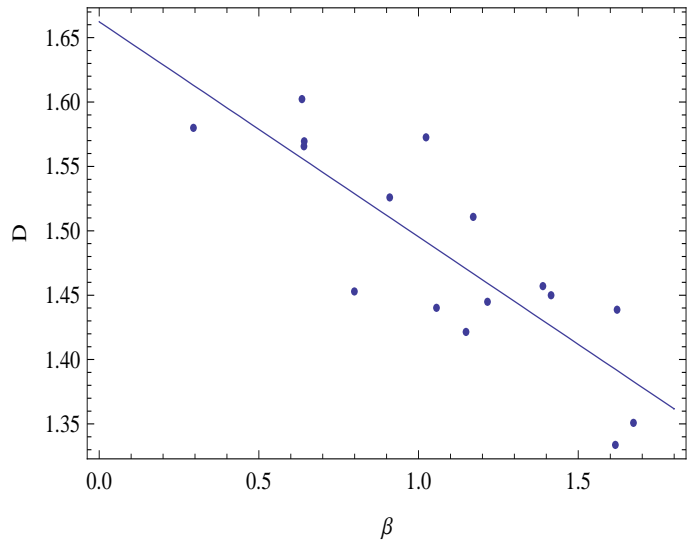}
\caption{Correlation between the fractal dimension $D$ and the spectral index $\beta $.}
\label{fig4}
\end{figure}

The relation between the theoretical data fit, given by Eq.~(\ref{eqc}) and the observational data is presented graphically in Fig.~\ref{fig4}.  The presence of the strong  Brownian noise in the IDV spectra also strongly support the idea that random stochastic processes play a significant role in their formation and evolution, as suggested in \cite{Leung}.

 \section{Discussions and final remarks}\label{sect4}

Since the determination of the fractal dimension of the times series  is relatively simple, it can provide useful information about the signal fluctuation from variable astrophysical sources. In the present paper we have explicitly obtained the fractal dimensions for several IDV emissions in both optical and radio bands. From the point of view of the fractal analysis the optical and the radio signals shows some remarkable differences. In the optical and X-ray bands the fractal dimension of the IDV signals is very close to the value $D=1.5$, indicating an almost perfect "Brownian noise" (random walk, or Wiener process) fluctuation spectrum (Brownian noise can be generated by integrating white noise) (\citealt{Hav, Den}). The predictability index of this signal is very low, showing that the emission in the optical band is dominated by purely stochastic processes. On the other hand the radio band emission pattern of the compact extragalactic radio source J 1128+5925 shows a relatively low fractal dimension of the order of $D=1.25$, with a much higher predictability index. This result can be interpreted from a physical point of view in the framework of the thermodynamic theory of fluctuations as follows. Assuming that in both fluctuating phases the self-gravitating fluctuations have a time dependence given by Eq.~(\ref{time}), we have $\alpha =\gamma =2/3$. Then from Eq.~(\ref{inter}) it follows that the time dependence of the fluctuation spectrum is of the order of $\sim t^{\alpha +\gamma }\sim t^{4/3}\sim t^{1.33}$, which gives a fractal dimension $D=1.33$, very close to the observed values of $D=1.25$ and $D=1.28$, respectively. Therefore this implies that the dominant time scale in the emission process is of the order of the gravitational free fall time scale. On the other hand, in order to obtain through the same method the fractal dimension ($D=1.5$) in the optical phase of the IDV the assumption of a $t^{\pm 0.75}$ dependence of the fluctuation spectrum in the fluctuating phases is necessary. This shows that the characteristic time scale of the density fluctuations cannot be of the same order of magnitude as the free fall time, and other physical processes, like, for example, viscous dissipation, may be responsible for the dynamics.

We have also obtained the general time dependence of the fluctuation spectrum in a self-gravitating system, as given by Eqs.~(\ref{eq1}) and (\ref{eq2}), respectively. The density fluctuations can be related to the luminosity fluctuations of the emitting source. By assuming that the emission of the radiation is in the form of thermal radiation, from the Stefan-Boltzmann law we find that the luminosity (energy emitted per second) from a spherical surface with radius $R$ is given by $L=4\pi R^2 \sigma T_{eff}^4$, where $\sigma $ is Stefan's constant, and $T_{eff}$ is the effective temperature of the emitting surface. With the use of the hydrostatic equilibrium equation and of the virial theorem one can show that $L\propto \rho ^{1/3}$ (\citealt{Choud}). With the use of this relation we obtain immediately $\Delta L/L\propto \Delta \rho /\rho $. Therefore, with the use of Eqs.~(\ref{eq1}) and (\ref{eq2}) the radial distance and time spectrum of the fluctuations of the luminosity in self-gravitating systems can be written as
\begin{equation}\label{eq28}
\sqrt{\left\langle \left(\frac{\Delta L }{L }\right)^2\right \rangle}\propto r^{-D_r }\propto t^{-D_t }.
\end{equation}

The constants $D_r$ and $D_t$ can be interpreted physically as the fractal dimensions, describing the fractal properties of the thermal radiation emission spectra from fluctuating sources.   The comparison of the fractal dimension of these spectra with the IDV observational data could give a powerful indication about the nature of the physical processes (thermodynamical or gravitational, random or deterministic) that play the dominant role in the electromagnetic emission process.

In the present paper we have also studied the correlation between the fractal dimension and the power spectral index $\beta $. We have found a very good correlation between these two physical parameters.

Fractal properties, or self-similarity, can be found in many aspects in nature. Our analysis has concentrated on the fractal properties of the IDV signal. Due to the limited amount of data used, our results can be considered only as a first step in the investigation of the fractal properties of the IDV's. It will be important in the future to investigate more signals at different wavelengths, in order to obtain a more accurate determination of the fractal dimension, and to clarify the relation between the fractal dimensions of different bands and the spectral power index, respectively.

\normalem
\begin{acknowledgements}
The work of T. H. was supported by a GRF grant of the Government of the Hong Kong SAR.
\end{acknowledgements}


\label{lastpage}

\section*{Appendix}

To obtain the spectral index $\beta $ and for fitting our data, we used the periodogram method as developed in \cite{V2010}.  The periodogram is used for dealing with the evenly sampled time series $x_k$ of $K$ points at intervals $\Delta T$. The definition of the periodogram is (\citealt{V05})
\begin{equation}
I(f_j)=\frac{2\Delta T}{ <x>^2N}|X_j|^2
\end{equation}
where $|X_j|^2$ is the modulus-squared of the discrete Fourier transform.

The unit of the normalization of the periodogram is $({\rm rms}/{\rm mean})^2\;{\rm Hz}^{-1}$. The initial purpose of the periodogram was to search for "hidden periodicity" in time series analysis. But, for a single time series, the periodogram of a noise process shows a great deal of scatter around the underlying power spectrum. For the case of a given frequency, $I\left(f_j\right)$ is scattered around the true power spectrum, $P\left(f_j\right)$. It also follows a $\chi ^2$ distribution with two degrees of freedom,
\begin{equation}\label{31}
I\left(f_j\right)=P\left(f_j\right)\chi ^2_2/2
\end{equation}
where  $\chi ^2_2$ is a random variable distributed as  $\chi^2$ with two degrees of freedom. It is equivalent with an exponential probability density, with a mean and variance of two and four, respectively,
\begin{equation}\label{32}
p_{\chi ^2}(x)=\exp\left(-x/2\right)/2.
\end{equation}

For the case of white noise flat spectrum, the power level is calculated by making use of Eqs.~(\ref{31}) and (\ref{32}) to obtain the maximum likelihood  of a given periodogram. Similarly, for the more general case of non-white noise, we need to examine the spurious peaks in the periodogram, and regard it as Brownian (red) noise. A simple method to obtain a reasonable estimate of power-law spectrum is $S(f)=\alpha f^{-\beta }$. $\alpha $ is the so called normalization of power-law power spectrum, and $\beta $ is the power spectral index.
\newpage


 \end{document}